\documentclass[aps,prl,preprintnumbers,showpacs,nofootinbib,floatfix]{revtex4}
\usepackage{graphicx}
\usepackage{comment}
\usepackage{dcolumn}
\usepackage{bm}
\begin{document}
\newcommand{\newc}{\newcommand}
\newc{\ra}{\rightarrow}
\newc{\lra}{\leftrightarrow}
\newc{\beq}{\begin{equation}}
\newc{\eeq}{\end{equation}}
\newc{\barr}{\begin{eqnarray}}
\newc{\earr}{\end{eqnarray}}
\newcommand{\Od}{{\cal O}}
\newcommand{\lsim}   {\mathrel{\mathop{\kern 0pt \rlap
  {\raise.2ex\hbox{$<$}}}
  \lower.9ex\hbox{\kern-.190em $\sim$}}}
\newcommand{\gsim}   {\mathrel{\mathop{\kern 0pt \rlap
  {\raise.2ex\hbox{$>$}}}
  \lower.9ex\hbox{\kern-.190em $\sim$}}}

\title{DIRECT WIMP DETECTION
IN DIRECTIONAL EXPERIMENTS
 }

\author{J. D. Vergados$^{(1),(2)}$\thanks{Vergados@cc.uoi.gr} and  Amand Faessler$^{(2)}$ }
\affiliation{$^{(1)}${\it Theoretical Physics Division, University
of Ioannina, Ioannina, Gr 451 10, Greece,}}
\affiliation{$^{(2)}${\it Institute of Theoretical Physics, University of Tuebingen, Tuebingen
Germany}}
\begin{abstract}
The recent WMAP data have confirmed that exotic dark matter
together with the vacuum energy (cosmological constant) dominate
in the flat Universe. Thus the direct dark matter search,
consisting of detecting the recoiling nucleus, is central to
particle physics and cosmology. Modern particle theories naturally provide viable cold dark matter
candidates with masses in the GeV-TeV region. Supersymmetry provides the lightest supersymmetric particle (LSP),
theories in extra
dimensions the lightest Kaluza-Klein particle (LKP) etc. In such theories the
 expected rates are much lower than the present experimental goals.
So one should exploit
characteristic signatures of the reaction, such as
the modulation effect and, in directional experiments, the
correlation of the event rates with the sun's motion.
 In standard non directional experiments the modulation is small,
less than two per cent and the location of the maximum depends on the unknown particle's mass.
 In directional experiments, in addition to the forward-backward asymmetry
due to the sun's motion, one expects a larger modulation, which depends on the direction
of observation. We study such effects both in the case of a light and a heavy target. Furthermore, since
it now appears that the planned experiments will be partly directional, in the sense that
they can only detect the line of the
recoiling nucleus, but not the sense of direction on it, we study which of the above mentioned interesting features, if any, will persist in these less ambitious experiments.
\end{abstract}
\date{\today}
\maketitle
\section{Introduction}
The combined MAXIMA-1 \cite{MAXIMA-1}, BOOMERANG \cite{BOOMERANG},
DASI \cite{DASI} and COBE/DMR Cosmic Microwave Background (CMB)
observations \cite{COBE} imply that the Universe is flat
\cite{flat01}
 and that most of the matter in
the Universe is Dark \cite{SPERGEL}. i.e. exotic. These results have been confirmed and improved
by the recent WMAP data \cite{WMAP06}. Combining the
the data of these quite precise experiments, crudely speaking, one finds:
$$\Omega_b=0.05, \Omega _{CDM}= 0.25, \Omega_{\Lambda}= 0.70$$
Since the non exotic component cannot exceed $40\%$ of the CDM
~\cite {Benne}, there is room for the exotic WIMP's (Weakly
Interacting Massive Particles).

 Even though there exists firm indirect evidence for a halo of dark matter
 in galaxies from the
 observed rotational curves, it is essential to directly
detect \cite{GOODWIT}-\cite{KVprd}
 such matter.
Until dark matter is actually detected, we shall not be able to
exclude the possibility that the rotation curves result from a
modification of the laws of nature as we currently view them.  This makes it imperative that we
invest a
maximum effort in attempting to detect dark matter whenever it is
possible. Furthermore such a direct detection will also
 unravel the nature of the constituents of dark matter.
 The
 possibility of such detection, however, depends on the nature of the dark matter
 constituents.

 Supersymmetry naturally provides candidates for the dark matter constituents
 \cite{GOODWIT}-\cite{ELLROSZ}.
 In the most favored scenario of supersymmetry the
LSP (Lightest Supersymmetric Particle) can be simply described as a Majorana fermion, a linear
combination of the neutral components of the gauginos and
higgsinos \cite{GOODWIT}-\cite{ref2}. In most calculations the
neutralino is assumed to be primarily a gaugino, usually a bino.

  Since the WIMP's are  expected to be very massive, $m_{WIMP} \geq 30 GeV$,  and
extremely non relativistic, with average kinetic energy $\prec T\succ  \approx
50KeV (m_{WIMP}/ 100 GeV)$, they are not likely to excite the nucleus.
So they can be directly detected mainly via the recoiling of a nucleus
(A,Z) in elastic scattering. The event rate for such a process can
be computed from the following ingredients:
\begin{enumerate}
\item An effective Lagrangian at the elementary particle (quark)
level obtained in the framework of the prevailing particle theory. For supersymmetry
this is achieved as described ,
e.g., in Refs~\cite{ref2,JDV96}.
\item A well defined procedure for transforming the amplitude
obtained using the previous effective Lagrangian from the quark to
the nucleon level, i.e. a quark model for the nucleon. This step
in SUSY models is not trivial, since the obtained results depend crucially on the
content of the nucleon in quarks other than u and d.
\item Knowledge of the relevant nuclear matrix elements
\cite{Ress}$-$\cite{DIVA00}, obtained with as reliable as possible many
body nuclear wave functions. Fortunately in the case of the scalar
coupling, which is viewed as the most important, the situation is
a bit simpler, since  then one  needs only the nuclear form
factor.
\item Knowledge of the WIMP density in our vicinity and its velocity distribution. Since the
essential input here comes from the rotational curves,  dark
matter candidates other than the LSP (neutralino) are also
characterized by similar parameters.
\end{enumerate}
 In the standard nuclear recoil experiments one has the problem that the reaction of interest does not have a characteristic feature to distinguish it
from the background. So for the expected low counting rates the background is
a formidable problem. Some special features of the LSP-nuclear interaction can be exploited to reduce the background problems. Such are:
\begin{itemize}
\item The modulation effect.\\
This yields e periodic signal due to the motion of the earth around the sun. Unfortunately this effect is small, $<2\%$ for most targets. Furthermore
it is inevitable to have backgrounds with a seasonal variation.
\item Transitions to excited states.\\
In this case one need not measure nuclear recoils, but the de-excitation $\gamma$ rays. This can happen only in vary special cases since the average WIMP energy is too low to excite the nucleus. It has, however, been found that in the special case of the target $^{127}$I such a process is feasible \cite{VQS04} with branching ratios around $5\%$.
\item Detection of electrons produced during the WIMP-nucleus collision.\\
It turns out, however, that this production peaks at very low energies. So only gaseous TPC detectors can reach the desired level of $100$eV. In such a case the number of electrons detected may exceed the number of recoils for a target with high $Z$ \cite{VE05},\cite{MVE05}.
\item Detection of hard X-rays produced when the inner shell holes are filled.\\
  It has been found \cite{EMV05} that in the previous mechanism inner shell electrons can be ejected. These holes can be filled by the Auger process or X-ray emission. For a target like Xe these X-rays are in the $30$keV region with the rate of about 0.1 per recoil for a WIMP mass of $100$ GeV.
\end{itemize}
In the present paper we will focus on the characteristic signatures of the WIMP nucleus
interaction, which will manifest themselves in directional recoil experiments, i.e. experiments
in which the direction of the recoiling nucleus is observed \cite{DRIFT},\cite{SHIMIZU03},\cite{KUDRY04},\cite{GREEN05},\cite{KRAUSS},\cite{KRAUSS01},\cite{JDV03},\cite{JDVSPIN04}. We will concentrate on the standard
Maxwell-Boltzmann (M-B)
distribution for the WIMPs of our galaxy and we will not be concerned with other non thermal distributions,
even though
they may yield stronger directional signals. Among those one should
mention the late infall of dark matter into the galaxy, i.e caustic rings
 \cite{SIKIVI1,SIKIVI2,Verg01,Green,Gelmini}, dark matter orbiting the
 Sun \cite{KRAUSS} and Sagittarius dark matter \cite{GREEN02}.

We will present our results in such
a fashion that they do not depend on the specific properties of the dark matter candidate, except that
the candidate is cold and massive, $m_{WIMP}\succeq 10$ GeV. So the only parameters which count is the reduced
mass, the nuclear form factor and the velocity distribution. So our results apply to all WIMPs.
  In a previous paper we have found that the observed rate is correlated
 with the direction of the sun's motion \cite{JDV03,JDVSPIN04}. On top of this one will observe a time dependent variation
of the rate due to the motion of the earth. Those features cannot be masked by any known background.
Unfortunately, however, even the most ambitious of the planned experiments are not expected soon to distinguish the
two possible senses along the line of nuclear recoil  \cite{SPOONER.PC}. Such experiments cannot, e.g., measure the backward-forward asymmetry. on this occasion we will extend our previous directional calculations \cite{JDV03}-\cite{JDVSPIN04} and explore, which characteristics, if any,
 of the previous calculation persist, if the rates for both senses of the nuclear recoil are summed up.
 \section{Rates}
Before computing the event rates for WIMP nucleus scattering we should discuss the kinematics. In the case of the WIMP-nucleus collision we find that the momentum transfer to the nucleus is given by
\beq
q=2 \mu_r \upsilon \cos{\theta}, \eeq
where $\theta$ is  the angle between the WIMP velocity and the momentum of the outgoing nucleus and
$\mu_r$ is the reduced mass of the system. Instead of the angle $\theta$ one introduces the energy $Q$ transferred
to the nucleus, $Q=\frac{q^2}{2 A m_p}$ ($A m_p$ is the nuclear mass). Thus
$$2 \sin{\theta} \cos{\theta}d\theta=-\frac{A m_p}{2 (\mu_r \upsilon)^2} dQ$$
Furthermore for a given energy transfer the velocity $\upsilon$ is constrained to be
\beq
\upsilon\succeq \upsilon_{min}~,~\upsilon_{min}= \sqrt{\frac{ Q A m_p}{2}}\frac{1}{\mu_r}
\eeq
We will find it it convenient to introduce instead of the energy transfer the dimensionless quantity $u$
\beq
u=\frac{1}{2}(qb)^2=A m_pQb^2\Rightarrow u=\frac{Q}{Q_0}~~,~~Q_{0}=\frac{1}{Am_p b^2}\simeq 4.1\times 10^{4}~A^{-4/3}~KeV
\label{u.1}
\eeq
where $b$ is the nuclear (harmonic oscillator) size parameter.

It is clear that for a given energy transfer the velocity is restricted from below. We have already mentioned that the velocity is bounded from above by the escape velocity. We thus get
\beq
a \sqrt{u}\leq y\leq n y_{esc}~,~a= \left[\sqrt{2}\mu_r b \upsilon_0 \right ]^{-1}
\eeq
with $n\geq 1$ (see below).
\beq
2 \sin{\theta} \cos{\theta} d \theta=-\frac{a^2}{y^2} dy,~y=\frac{\upsilon}{\upsilon_0}
\eeq
The differential (non directional) rate with respect to the energy
transfer u can be written as:
\begin{equation}
dR_{undir} = \frac{\rho (0)}{m_{\chi}} \frac{m}{A m_N}
 d\sigma (u,\upsilon) | {\boldmath \upsilon}|
\label{2.18}
\end{equation}
 Where   $\rho (0) = 0.3 GeV/cm^3$ is the LSP density in our vicinity,
 m is the detector mass, $m_{\chi}$ is the WIMP  mass and
$d\sigma(u,\upsilon )$ the nucleus WIMP cross section.\\
 The corresponding directional differential rate, i.e. when only recoiling nuclei
 with non zero velocity in the direction $\hat{e}$  are observed, is given by :
\begin{eqnarray}
dR_{dir} &=& \frac{\rho (0)}{m_{\chi}} \frac{m}{A m_N}
|\upsilon| \hat{\upsilon}.\hat{e} ~\Theta(\hat{\upsilon}.\hat{e})
 ~\frac{1}{2 \pi}~
d\sigma (u,\upsilon)\\
\nonumber & &\delta(\frac{\sqrt{u}}{\mu_r \upsilon
b\sqrt{2}}-\hat{\upsilon}.\hat{e})
 ~~,~ \Theta (x)= \left \{
\begin{array}{c}1~,x>0\\0~,x<0 \end{array} \right \}
 \label{2.20}
\end{eqnarray}

The LSP is characterized by a velocity distribution. For a given
velocity distribution f(\mbox{\boldmath $\upsilon$}$^{\prime}$),
 with respect to the center of the galaxy,
One can find the  velocity distribution in the lab frame
$f(\mbox{\boldmath $\upsilon$},\mbox{\boldmath $\upsilon$}_E)$
by writing

\hspace{2.0cm}\mbox{\boldmath $\upsilon$}$^{'}$=
          \mbox{\boldmath $\upsilon$}$ \, + \,$ \mbox{\boldmath $\upsilon$}$_E
 \,$ ,
\hspace{2.0cm}\mbox{\boldmath $\upsilon$}$_E$=\mbox{\boldmath $\upsilon$}$_0$+
 \mbox{\boldmath $\upsilon$}$_1$

\mbox{\boldmath $\upsilon$}$_0 \,$  is the sun's velocity (around
the center of the galaxy), which coincides with the parameter of
the Maxwellian distribution, and \mbox{\boldmath $\upsilon$}$_1
\,$ the Earth's velocity
 (around the sun).
 The velocity of the earth is given by
\begin{equation}
\mbox{\boldmath $\upsilon$}_E  = \mbox{\boldmath $\upsilon$}_0 \hat{z} +
                                  \mbox{\boldmath $\upsilon$}_1
(\, sin{\alpha} \, {\bf \hat x}
-cos {\alpha} \, cos{\gamma} \, {\bf \hat y}
+ cos {\alpha} \, sin{\gamma} \, {\bf \hat z} \,)
\label{3.6}
\end{equation}
In the above formula $\hat{z}$ is in the direction of the sun`s motion,
$\hat{x}$ is in the radial direction out of the galaxy,  $\hat{y}$ is
perpendicular in the plane of the galaxy ($\hat{y}=\hat{z} \times \hat{x}$)
and $\gamma \approx \pi /6$ is the inclination of the axis of the ecliptic
 with respect to the plane of the galaxy. $\alpha$ is the phase of the Earth
in its motion around the sun ($\alpha=0$ around June 2nd).

The above expressions for the rates must be folded with the LSP velocity
 distribution. In the present work we will assume that the velocity distribution is Maxwell-Boltzmann (M-B) with a characteristic
 velocity $\upsilon_0$ and an upper cut off equal to the escape velocity, $\upsilon_{esc}=2.84 \upsilon_0$. For comparison we
 will also consider  M-B distributions with larger characteristic velocity
 $n \upsilon_0$ and cut off velocity  $2.84 n \upsilon_0$,
with $n\geq 1$. This situation arises, if one considers the coupling of dark matter to dark energy via a scalar field. Then the gravitational interaction of matter
is not affected. The interaction involving dark matter is increased. Via the virial theorem this results to an isothermal M-B distribution with higher temperature.
We will not elaborate further on this point, but we refer the interested reader to the literature. \cite{TETRADIS,TETRVER06}.

We will distinguish two possibilities:

\subsection{The direction of the recoiling nucleus is not observed.}
 Even though our main interest is in the directional rate for orientation purposes
 we will summarize the main points entering the standard (non directional) searches.
 The non-directional differential rate folded with the WIMP velocity distribution is given by:
\begin{equation}
\Big<\frac{dR_{undir}}{du}\Big> = \Big<\frac{dR}{du}\Big> =
\frac{\rho (0)}{m_{\chi}} \frac{m}{Am_N} \sqrt{\langle
\upsilon^2\rangle }\int
           \frac{   |{\boldmath \upsilon}|}
{\sqrt{ \langle \upsilon^2 \rangle}}
 f(\mbox{\boldmath $\upsilon$},\mbox{\boldmath $\upsilon$}_E)
                       \frac{d\sigma (u,\upsilon )}{du} d^3
 \mbox{\boldmath $\upsilon$}
\label{3.11}
\end{equation}
where
$$f(\mbox{\boldmath $\upsilon$},\mbox{\boldmath $\upsilon$}_E)\Rightarrow f(y,\xi,\phi,\alpha,n)$$
\barr
&&f(y,\xi,\phi,\alpha,n)=\frac{1}{\pi  \sqrt{\pi}} \frac{1}{n^3} e^X
\nonumber\\
X&=& \left(\frac{-y^2-2 \left(\delta ^2+y \sqrt{1-\xi ^2} \cos {\phi } \sin {\alpha )}\delta +\cos
   {\alpha } \left((y \xi +2) \sin {\gamma }-y \sqrt{1-\xi ^2} \cos {\gamma } \sin {\phi
   }\right) \delta +y \xi +1\right)}{n^2} \right)
   \nonumber\\
   &&
   \label{vdis}
\earr

The event rate for the coherent WIMP-nucleus elastic scattering is given by \cite{Verg01,JDV03,JDVSPIN04,JDV06}:
\beq
R= \frac{\rho (0)}{m_{\chi^0}} \frac{m}{m_p}~
              \sqrt{\langle v^2 \rangle } \left [f_{coh}(A,\mu_r(A)) \sigma_{p,\chi^0}^{S}+f_{spin}(A,\mu_r(A))\sigma _{p,\chi^0}^{spin}~\zeta_{spin} \right]
\label{fullrate}
\eeq
with
\beq
f_{coh}(A, \mu_r(A))=\frac{100\mbox{GeV}}{m_{\chi^0}}\left[ \frac{\mu_r(A)}{\mu_r(p)} \right]^2 A~t_{coh}\left(1+h_{coh}cos\alpha \right)
\eeq
\beq
f_{spin}(A, \mu_r(A))=\left[ \frac{\mu_r(A)}{\mu_r(p)} \right]^2 \frac{t_{spin}(A)}{A}
\eeq
with $\sigma_{p,\chi^0}^{S}$ and $\sigma _{p,\chi^0}^{spin}$ the scalar and spin proton cross sections
$~\zeta_{spin}$ the nuclear spin ME. In this work we will not be concerned with the spin cross section.

 The number of events in time $t$ due to the scalar interaction, which leads to coherence, is:
\beq
 R\simeq 1.60~10^{-3}
\frac{t}{1 \mbox{y}} \frac{\rho(0)}{ {\mbox0.3GeVcm^{-3}}}
\frac{m}{\mbox{1Kg}}\frac{ \sqrt{\langle
v^2 \rangle }}{280 {\mbox kms^{-1}}}\frac{\sigma_{p,\chi^0}^{S}}{10^{-6} \mbox{ pb}} f_{coh}(A, \mu_r(A))
\label{eventrate}
\eeq
In the above expression
 $m$ is the target mass, $A$ is the number of nucleons
in the nucleus and $\langle v^2 \rangle$ is the average value of the square of the WIMP velocity (with $n=1$).
The quantity of interest to us is the quantity $r= t_{coh}\left(1+h_{coh}cos\alpha \right)$, which contains all the information
regarding the WIMP velocity distribution and the structure of the nucleus. It also depends on the reduced mass of the system.
It is not difficult to show  \cite{Verg01,JDV03,JDVSPIN04,JDV06} that:

\begin{equation}
\frac{dr}{du}=\sqrt{\frac{2}{3}} a^2 F^2(u)  \Psi(a \sqrt{u},\alpha )
\label{eq:rrateall}
\end{equation}
where \beq \Psi(x,\alpha)=\int_{x}^{yesc}  y  dy
\int_{-1}^{\xi_0(y,\alpha )} d \xi \int_0^{2 \pi} d \phi
f(x,y,\xi,\phi,n) \eeq where $F(u)$ is the nuclear form factor.
The quantity $\xi_0(y,\alpha )$ enters since in some region of the
velocity space the upper value of $\xi$ is restricted so that the
condition \beq \sqrt{y^2+2 y \xi (1+ \delta \cos{\alpha}) +2(1+ 2
\delta \cos{\alpha}+ 4 \delta^2)}\preceq n y_{esc}
\label{condition} \eeq
 is satisfied.

 The $\phi$ integration can be performed to yield:
 \beq
\Psi(x,\alpha)=\int_{x}^{yesc}  y  dy \int_{-1}^{\xi_0(y,\alpha )} d \xi
f_1(x,y,\xi,n)
\eeq
with
 \barr
 f_1(y,\xi,\alpha,n)&=&\frac{2}{  \sqrt{\pi}} \frac{1}{n^3}Exp \left(\frac{-y^2-2 \left(\delta ^2+\cos
   {\alpha } \left((y \xi +2) \sin {\gamma }\right) \delta +y \xi +1\right)}{n^2} \right)
\nonumber\\
 & &I_0(2\delta y \sqrt{(1-\xi^2)(1-\cos^2{\alpha}\sin^2{\gamma})})
   \label{vdis1}
   \earr
   where $I_0$ is the well known modified Bessel function.

By performing a Fourier analysis of the function $\Psi(x,\alpha)$, which is a periodic function of $\alpha$,
 and keeping the dominant terms we obtain the two amplitudes $\Psi_0(a \sqrt{u})$ and  $H(a \sqrt{u})$. Thus:
 \beq
\frac{dr}{du}= \frac{dt}{du}+\frac{dh}{du}\cos{\alpha}
\eeq
where
\beq
\frac{dt}{du}=\sqrt{\frac{2}{3}} a^2 F^2(u) \Psi_0(a \sqrt{u})~~,~~\frac{dh}{du}=\sqrt{\frac{2}{3}} a^2 F^2(u)
H(a \sqrt{u})
\eeq
\\The total (time averaged) rate is given by:
 \beq
 t_{coh}=\int_{u_{min}}^{u_{max}} \frac{dt_{coh}}{du} du
 \label{tcoh}
 \eeq
 where
$$u_{min}\Leftrightarrow \mbox{detector threshold}~,~
u_{max}=\frac{(n y_{esc})^2}{a^2}\Leftrightarrow \mbox{maximum WIMP velocity}$$
By including  both $\Psi_0(a \sqrt{u})$ and $H(a \sqrt{u})$ we can cast the rate in the form:
\beq
r=t_{coh} \left(1+h_{coh} \cos{\alpha} \right)~,~h_{coh}=\frac{1}{t_{coh}} \int_{u_{min}}^{u_{max}} \frac{dh_{coh}}{du} du
\eeq
\subsection{ The direction $\hat{e}$ of the recoiling nucleus is observed.}
In this case the directional differential rate is given by:
\begin{eqnarray}
\Big<\frac{dR_{dir}}{du}\Big> =
\frac{\rho (0)}{m_{\chi}} \frac{m}{Am_N} \sqrt{\langle
\upsilon^2\rangle } \int \frac{
\mbox{\boldmath $\upsilon$}.\hat{e}~
            \Theta( \mbox{\boldmath $\upsilon$}.\hat{e})}
{\sqrt{ \langle \upsilon^2 \rangle}}
 f(\mbox{\boldmath $\upsilon$},\mbox{\boldmath $\upsilon$}_E)
                       \frac{d\sigma (u,\upsilon )}{du}\frac{1}{2 \pi}
 \delta(\frac{\sqrt{u}}{\mu_r b \upsilon}-\hat{\upsilon}.\hat{e}) d^3
 \mbox{\boldmath $\upsilon$}
\label{3.12b}
\end{eqnarray}
The factor of $1/ 2 \pi$ appears, since we are using the same cross section as in the non directional
case, even though no angular integration is now required.
 The above coordinate system, properly taking into account the motion of
 the sun
and the geometry of the galaxy, is not the most convenient for performing
the needed integrations in the case of the directional expressions.
 For this purpose we go to another coordinate
system in which the polar axis, $\hat{Z}$, is in the direction of
observation (direction of the recoiling nucleus) via the
transformation:
$$\left ( \begin{array}{c}\hat{X} \\
 \hat{Y} \\
\hat{Z} \end{array} \right )=
\left ( \begin{array}{ccc}
\cos{\Theta}\cos{\Phi}&\cos{\Theta}\sin{\Phi}&-\sin{\Theta}\\
-sin{\Phi}&  \cos{\Phi}& 0\\
\sin{\Theta}\cos{\Phi}&\sin{\Theta}\sin{\Phi}&\cos{\Theta}\\
 \end{array} \right )
\left ( \begin{array}{c}\hat{x} \\
\hat{y} \\
\hat{z}
 \label{transf}
 \end{array} \right )$$
In this coordinate system the orientation parameters $\Theta$
and $\Phi$ appear  explicitly in the distribution function. In fact the numerator of the exponent of the M-B distribution
 (\ref{vdis}) becomes:
$$
\delta ^2+2 y \xi  \cos {\Phi} \sin {\alpha } \sin {\Theta}
   \delta -2 y \xi  \cos {\alpha } \cos {\gamma } \sin {\Phi} \sin
   {\Theta} \delta +2 \cos {\alpha } \sin {\gamma } \delta
 -2 y
   \sqrt{1-\xi ^2} \cos {\alpha } \cos {\Phi} \cos {\gamma } \sin
   (\phi ) \delta+
   $$
   $$
-2 y \sqrt{1-\xi ^2} \sin {\alpha } \sin {\Phi}
   \sin {\phi } \delta +y^2-2 y \sqrt{1-\xi ^2} \cos {\phi } \sin
   {\Theta} (\delta  \cos {\alpha } \sin {\gamma }+1)+$$
$$+2 y \cos
   {\Theta} \left(\delta  \cos {\alpha } \sin {\gamma } \xi +\xi
   +\delta  \sqrt{1-\xi ^2} \cos {\Phi} \cos {\phi } \sin {\alpha
   }-\delta  \sqrt{1-\xi ^2} \cos {\alpha } \cos {\gamma } \cos (\phi )
   \sin {\Phi}\right)+1$$
This is clearly quite messy, but the constraint in the integration variables imposed by the Heaviside function becomes
quite simple, namely the velocity in polar coordinates is specified by the angles $ 0\leq  \theta\leq \pi/2 $ , $ 0\leq \phi\leq 2 \pi $.

 The $\delta$ function ensures
that in the directional case the variables $u,\upsilon$ and $\xi$
obey the required relation.
In our numerical calculation we found
it more convenient to use $y=\upsilon/\upsilon_0$ and to express
$\xi$ in terms of the other two, namely $u$ and $y$. So one is
left with two integrations, over $\phi$ and $y$. To  make the calculations tractable we made an expansion to first order
in $\delta=\frac{\upsilon_1}{\upsilon_0}=0.135$. Thus the distribution takes the form:
\begin{eqnarray}
& &f(\mbox{\boldmath $\upsilon$},\mbox{\boldmath $\upsilon$}_E,\Theta,\Phi)=\frac{1}{m^5 \pi ^{3/2}}
Exp \left( {-\frac{y^2-2 \sqrt{y^2-x^2} \cos {\phi } \sin
   {\Theta}+2 x \cos {\Theta}+1}{m^2}} \right )
\nonumber\\
&&[ m^2+
 +2 \delta
   \cos {\alpha } x \cos {\gamma } \sin {\Phi} \sin
   {\Theta}+\left(\sqrt{y^2-x^2}  \cos {\phi } \sin
   {\Theta}-1\right) \sin {\gamma }
   \nonumber\\
& &+2 \delta
   \cos {\alpha }\cos {\Theta}
   \left(\sqrt{y^2-x^2}  \cos {\gamma } \cos {\phi } \sin
   {\Phi}-x \sin (\gamma )\right) \delta
   \sqrt{y^2-x^2}  \sin {\alpha } \sin {\Phi} \sin {\phi
   }
\nonumber\\
&&-2 \delta  \cos {\Phi} \left(\sqrt{y^2-x^2}  \cos
   {\Theta} \cos {\phi } \sin {\alpha }+x \sin {\Theta} \sin
   {\alpha }-\sqrt{y^2-x^2}  \cos {\alpha } \cos {\gamma } \sin
   {\phi }\right)]
\end{eqnarray}
The integration over the angle $\phi$ can now easily be accomplished yielding:
\begin{itemize}
\item Time independent part:
\beq
f^{dir}_0(x,y,\Theta)=\frac{2}{m^3 \sqrt{\pi}}Exp \left( {-\frac{y^2+2 x \cos {\Theta}+1}{m^2}} \right) I_0(\frac{2\sqrt{y^2-x^2}\sin{\Theta}}{m^2})
\label{fdir0}
\eeq
 Note that this part is independent of $\Phi$.
\item The modulation amplitude proportional to $\cos{\alpha}$
\begin{eqnarray}
f^{dir}_c(x,y,\Theta,\Phi)&=&\frac{4}{m^5 \sqrt{\pi }} \delta  Exp \left({-\frac{y^2+2 x \cos
   {\Theta}+1}{m^2}} \right)
\\
\nonumber
&& [I_0\left(\frac{2
   \sqrt{y^2-x^2} \sin {\Theta}}{m^2}\right) (x
   \cos {\gamma } \sin {\Phi} \sin
   {\Theta}-(x \cos {\Theta}+1) \sin
   {\gamma })
\\
\nonumber
&+&\sqrt{y^2-x^2} I_1\left(\frac{2
   \sqrt{y^2-x^2} \sin {\Theta}}{m^2}\right)
   (\cos {\Theta} \cos {\gamma } \sin
   {\Phi}+\sin {\Theta} \sin {\gamma
   })]
\end{eqnarray}
\item The modulation amplitude which is proportional to $\sin{\alpha}$:
\begin{eqnarray}
f^{dir}_s(x,y,\Theta,\Phi)&=&
-\frac{4}{m^5 \sqrt{\pi }} \delta  Exp \left({-\frac{y^2+2 x \cos
   {\Theta}+1}{m^2}} \right) \cos {\Phi}
   \\
   \nonumber
   &&\left(\sqrt{y^2-x^2}  I_1\left(\frac{2
   \sqrt{y^2-x^2}  \sin
   {\Theta}}{m^2}\right) \cos {\Theta}+x
   I_0\left(\frac{2 \sqrt{y^2-x^2}  \sin
   {\Theta}}{m^2}\right) \sin
   {\Theta}\right)
   \end{eqnarray}
   Such an amplitude does not appear to leading order in non directional rate. Here it may become important
   near $\Phi=0$  or $\pi$.
\end{itemize}
In the above expressions  $I_0$ and $I_1$ are the modified Bessel functions.

From now on the needed integrations over $y$ and $u$ can be done numerically. Thus one obtains:
\beq
\Psi^{dir}_i(x,\Theta,\Phi)=\int_x^{yesc} y f^{dir}_i(x,y,\Theta,\Phi) dy,~i=0,c,s;
\eeq
\beq H_m((x,\Theta,\Phi))=\sqrt{(\Psi^{dir}_c(x,\Theta,\Phi))^2+
(\Psi^{dir}_s(x,\Theta,\Phi))^2 }
\eeq
The last function maybe used in obtaining the magnitude of the modulation amplitude.
From now on we proceed as in the previous section, except that the obtained results are
functions of $\Theta$ and $\Phi$.
\begin{equation}
\left(\frac{dt}{du}\right)_{dir}= \sqrt{\frac{2}{3}} a^2 F^2(u)\Psi^{dir}_0(a \sqrt{u}),\left(\frac{dh}{du}\right)_{dir}=\sqrt{\frac{2}{3}} a^2 F^2(u)\left[ \Psi^{dir}_c(a \sqrt{u}) \cos{\alpha}+\Psi^{dir}_s(a \sqrt{u}) \sin{ \alpha} \right ]
\label{eq:drrate1}
\end{equation}
or equivalently
\begin{equation}
\left(\frac{dh}{du}\right)_{dir}= \sqrt{\frac{2}{3}} a^2 F^2(u) H_m(a \sqrt{u}) \cos{(\alpha+\vartheta  _m \pi)}
\label{eq:drrate2}
\end{equation}
where $\vartheta _m$ is the shift in the phase of the modulation (in units of $\pi$) relative to the phase of the Earth, namely:
 \beq
 \vartheta _m=-\frac{1}{\pi}\arctan{\frac{\Psi^{dir}_s(a \sqrt{u})}{\Psi^{dir}_c(a \sqrt{u})}}
\eeq
The event rate is still given by Eq. (\ref{eventrate}) except that now:
 \beq
f_{coh}(A, \mu_r(A))=\frac{100\mbox{GeV}}{m_{\chi^0}}\left[ \frac{\mu_r(A)}{\mu_r(p)} \right]^2 A\frac{\kappa}{2 \pi}t_{coh}\left(1+h_m(coh)cos{(\alpha+\alpha_m \pi)} \right)
\eeq
with
\beq
\kappa=\frac{t_{dir,coh}}{t_{coh}}
\eeq
and
 \beq
 t_{dir,coh}=\int_{u_{min}}^{u_{max}} \left(\frac{dt_{coh}}{du} \right)_{dir} du
 \label{dtcoh}
 \eeq
$\kappa$ is a measure of the reduction in the event rate in directional experiments over and above the geometric factor of $1/(2 \pi)$. Furthermore
 \beq
 h_c \cos{\alpha}+h_s \sin{\alpha}=\frac{1}{t^{dir}(coh)}\int_{u_{min}}^{u_{max}} \left(\frac{dh_{coh}}{du} \right)_{dir} du
 \label{dhcoh}
 \eeq
 \beq
 h_m(coh)=\sqrt{h_c^2+h_s^2},~\alpha_m=-\frac{1}{\pi}\arctan{\frac{h_s}{h_c}}
 \eeq
It is clear that the quantifies $h_c$ and $h_s$ in Eq. (\ref{dhcoh}) are obtained from  $\Psi_c$
 and $\Psi_s$ respectively. Clearly the quantities $\kappa$, $h_m$ and
 $\alpha_m$ depend on the direction of observation. The range of the
 relevant angles needed to specify the line of recoil can be
chosen to be :
$$0\leq \Theta\leq \pi,~0\leq \Phi\leq 2 \pi$$
Sometimes in directional experiments one is interested in the
angular distribution of events around the direction of
observation, since the direction of the track may not be known
precisely \cite{DRIFT,GREEN02,GREEN01,GREEN05}. To get an expression for this, we proceed as
above by eliminating the variable $y$ in terms of $u$ and $\xi$.
Then for a given $\xi$ we integrate over the energy transfer u. In
this instance we will ignore the dependence of the event rate on
$\delta$, i.e.  we will neglect the modulation. Thus we define the
angular distribution of the expected events in time as follows: \beq
 \frac{dR_{dir,coh}}{d \xi}\simeq 1.60~10^{-3}
\frac{t}{1 \mbox{y}} \frac{\rho(0)}{ {\mbox0.3GeVcm^{-3}}}
\frac{m}{\mbox{1Kg}}\frac{ \sqrt{\langle v^2 \rangle }}{280 {\mbox
kms^{-1}}}\frac{\sigma_{p,\chi^0}^{S}}{10^{-6} \mbox{ pb}}
\frac{df_{dir,coh}(A, \mu_r(A))}{d \xi} \label{ksieventrate} \eeq
Ignoring the modulation effect we get: \beq \frac{df_{dir,coh}(A,
\mu_r(A))}{d \xi}=\frac{100\mbox{GeV}}{m_{\chi^0}}\left[
\frac{\mu_r(A)}{\mu_r(p)} \right]^2 \frac{A}{2 \pi}~ \frac{dr_{dir,coh}}{d \xi}
\eeq
\subsection{Partly directional experiments.}
In this case one can specify the line the nucleus is recoiling but not the sense of direction on it. The results in this case
are obtained from those of the previous section via the replacements of the functions defined above by:
\beq
\bar{f}^{dir}_0(x,y,\Theta)=f^{dir}_0(x,y,\Theta)+f^{dir}_0(x,y,\pi-\Theta)
\eeq
\beq
\bar{f}^{dir}_i(x,y,\Theta,\Phi)=f^{dir}_i(x,y,\Theta,\Phi)+f^{dir}_i(x,y,\pi-\Theta,\Phi+\pi),~i=c,s
\eeq
The range of the angles now is:
$$0\leq \Theta\leq \pi/2,~0\leq \Phi\leq \pi$$
\subsection{Some general results}
\begin{figure}[!ht]
 \begin{center}
 \includegraphics[scale=0.8]{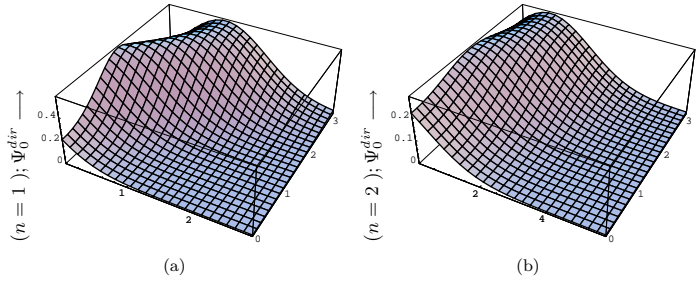}
 \caption{The  function $\Psi^{dir}_0$ as a function of $x$  and $\Theta$ (y-axis in radians). (a) $n=1$ and (b) $n=2$.
}
 \label{fig:3DT0}
  \end{center}
  \end{figure}

Before proceeding to specific applications involving specific nuclear targets it is instructive to make
general observations, ignoring the fact that eventually $x=a \sqrt{u}$, with $a$ depending, among other
things, on the specific target.
First we note that the time average rate depends only on the polar angle $\Theta$ (see Fig.
\ref{fig:3DT0}).
From Fig. \ref{fig:3DT0} we see that for large $x$ the function $\Psi^{dir}_0$ peaks in the direction opposite
to the sun's
direction of motion , as expected for a M-B distribution. This, however is not true for small $x$. This is
understood by observing that $x$ appears explicitly as  the coefficient of $\cos{\Theta}$
in the exponential of Eq. (\ref{fdir0} )
 The situation for the modulated amplitude $H_m$ is more complicated since it also depends on the angle $\Phi$.
Therefore we will not discuss this case here. We only mention  that
the modulation here in the case of the $n=2$
  is  suppressed relative to that for the standard $n=1$ case,
 independently of the angle of observation.
 This is expected in view of the results for the
  standard (non directional) case \cite{TETRVER06}.
  \section{Some applications}
In this section we are going to apply the formalism of the previous section in
 the case of two popular targets: i) The light target  $^{32}S$ appearing in CS$_2$ involved in DRIFT \cite{DRIFT}
  and ii) The
$^{127}$I target, which has been  employed in the DAMA experiment \cite{BERNA1,BERNA2}. We will not consider energy 
thresh hold effects and we will
ignore quenching factor effects. We will consider only the coherent mode, but the results obtained for the functions
$t,\kappa,h_m$ and $\alpha_m$ are not expected to be radically modified, if one considers the spin mode.
\subsection{The light target CS$_2$}
\begin{figure}[!ht]
 \begin{center}
  \includegraphics[scale=0.8]{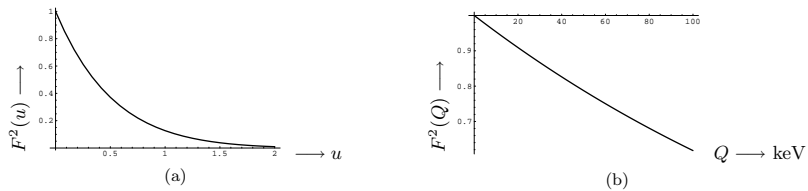}
   \caption{(a) The form factor $F^2(u)$ for $^{32}$S employed in our calculation with  $u=Q/Q_0$, Q  the
 energy transfer to the nucleus   and $Q_0=404$ keV. (b) The same quantity a function of the energy transfer $Q$.}
  \label{fig:sqformf32}
   \end{center}
  \end{figure}

\begin{figure}[!ht]
 \begin{center}
   \includegraphics[scale=0.8]{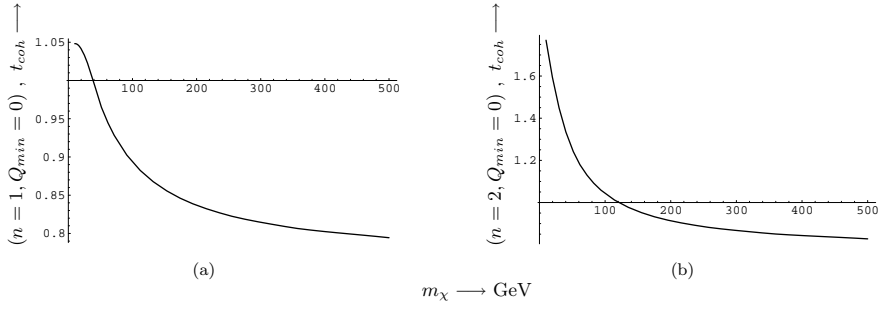}
 \caption{The quantity $t_{coh}$ is shown for  $Q_{min}=0$ with (a) $n=1$  and (b) $n=2$.}
 \label{fig:Stotalt}
   \end{center}
  \end{figure}

The nuclear form factor employed was obtained in the shell model description of the target and is shown in
 Fig. \ref{fig:sqformf32}.
  The parameter $t$ entering the non directional case is shown in Fig. \ref{fig:Stotalt}

\begin{figure}[!ht]
 \begin{center}
   \includegraphics[scale=0.8]{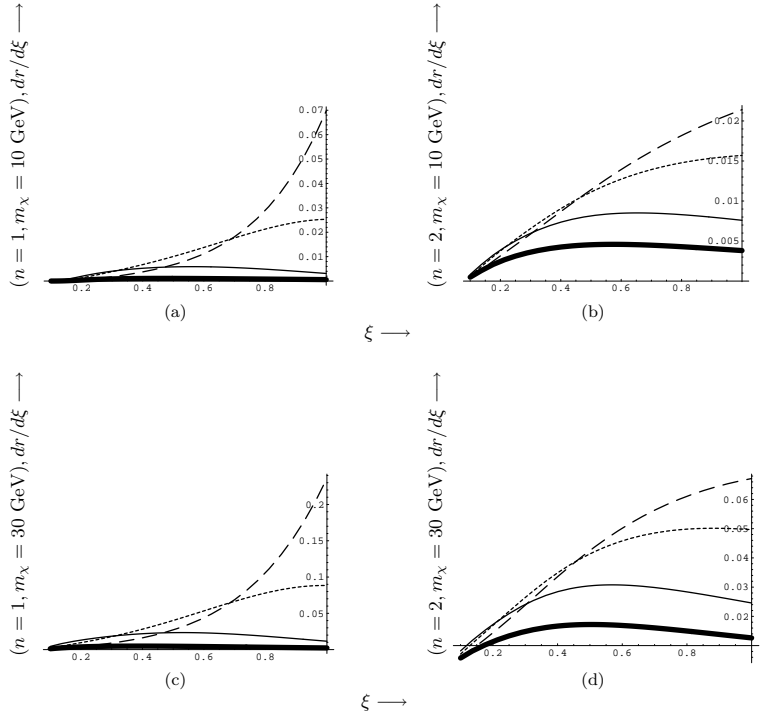}
\caption{The differential rate $dr/d \xi$ for $^{32}$S in  as function of $\xi$, the cosine of the angle between
  the line of observation and the line of recoil. The thick solid, fine solid, short,  and long   dash correspond
  to $\Theta=\pi/4,\pi/2, 3\pi/4$ and $\pi$ respectively. (a) directional in a given sense for $n=1$ in the case
   of a WIMP mass
of $10$ GeV. (b) The same as in (a) for $n=2$. (c) the same as in (a) for 30 GeV. (d) the same as in (b) for
 30 GeV. }
 \label{fig:dRdksi1}
  \end{center}
  \end{figure}

\begin{figure}[!ht]
 \begin{center}
    \includegraphics[scale=0.8]{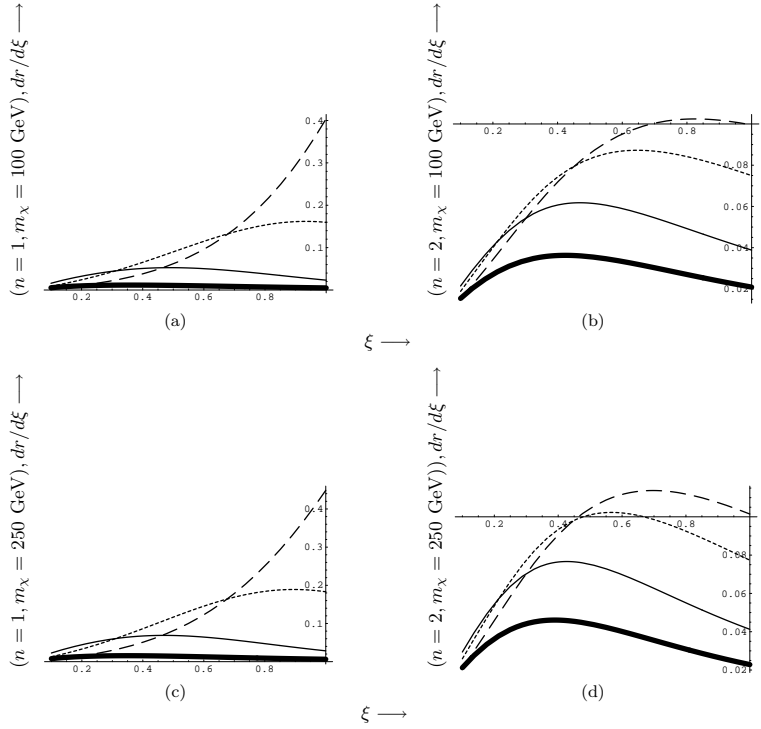}
 \caption{The same as in Fig. \ref{fig:dRdksi1}. (a) directional in a given sense for $n=1$ for $100$ GeV. (b) The same as in (a) for $n=2$. (c) the same as in (a) for 250 GeV. (d) the same as in (b) for
 250 GeV.}
 \label{fig:dRdksi2}
  \end{center}
  \end{figure}

\begin{figure}[!ht]
 \begin{center}
     \includegraphics[scale=0.8]{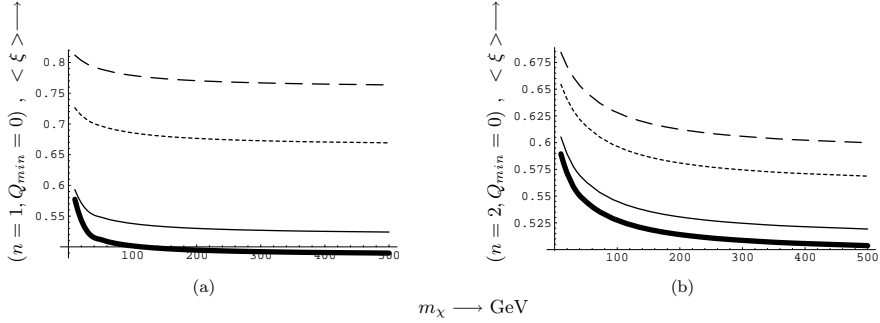}
  \caption{The quantity $<\xi>$ is shown for $Q_{min}=0$ with (a) $n=1$  and (b) $n=2$. Otherwise the
  notation is
that of Fig. \ref{fig:dRdksi1}}
 \label{fig:aveksi}
   \end{center}
  \end{figure}

 We begin our analysis of the directional signal by computing the differential rate with respect to $\xi$,
 which is used in simulating the experiments \cite{DRIFT,GREEN02,GREEN01,GREEN05}.
 Our results are presented in Fig. \ref{fig:dRdksi1} and \ref{fig:dRdksi2}.
With the above angular distribution we can obtain the average
value of $\xi$, $<\xi>$. Thus we get the results shown in Fig.
\ref{fig:aveksi}.
  We see that the average value of $\xi$ does not change vary much
in going from the $n=1$ to the $n=2$ case. We clearly see that
this value is much higher in the case the observation is made
opposite to the sun's direction of motion ($\Theta=\pi$).

\begin{figure}[!ht]
 \begin{center}
    \includegraphics[scale=0.8]{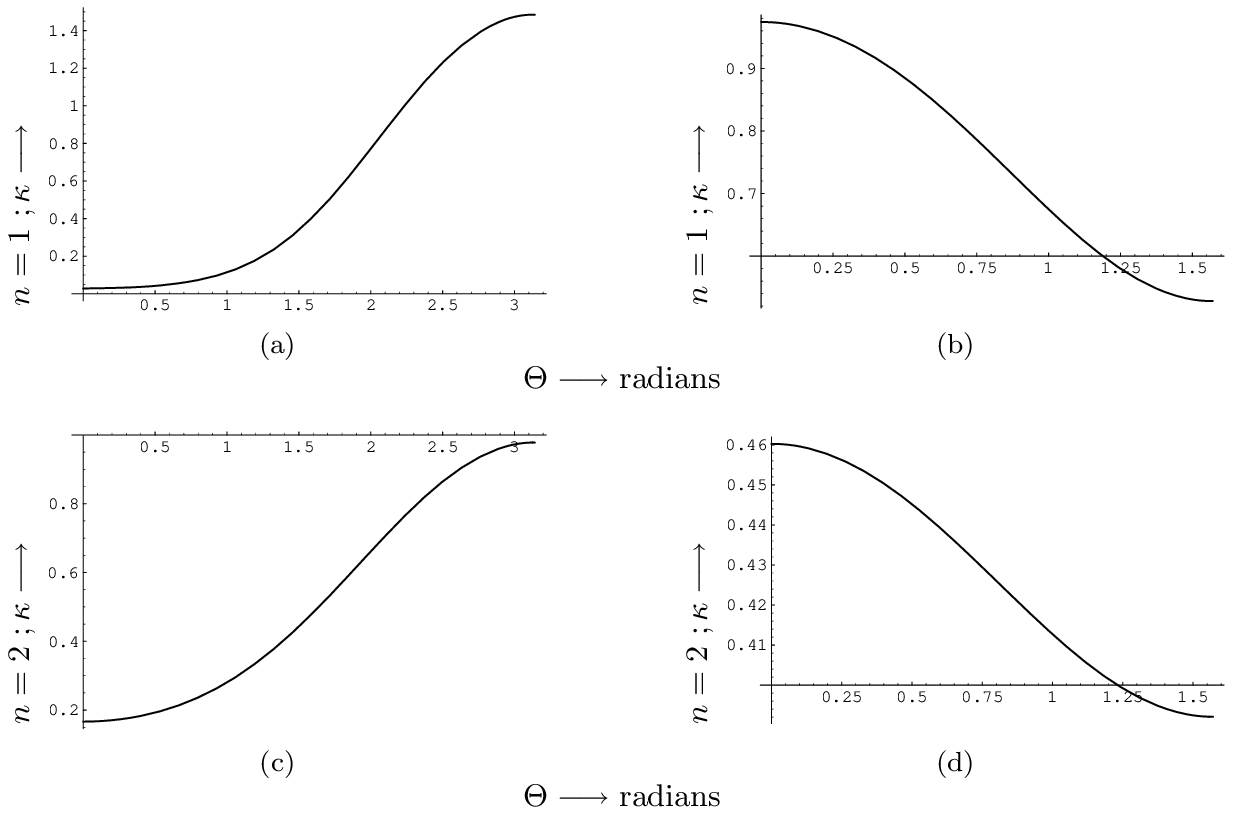}
 \caption{The parameter $\kappa$ defined in the text for a light target in the case of  a WIMP mass
of $10$ GeV. (a) Directional in a given sense for $n=1$. (b) The
sum of both senses for $n=1$. (c) The same as in (a) for $n=2$.
(d) The same as in (b) for $n=2$.}
 \label{fig:Skappa1}
  \end{center}
  \end{figure}

\begin{figure}[!ht]
 \begin{center}
   \includegraphics[scale=0.8]{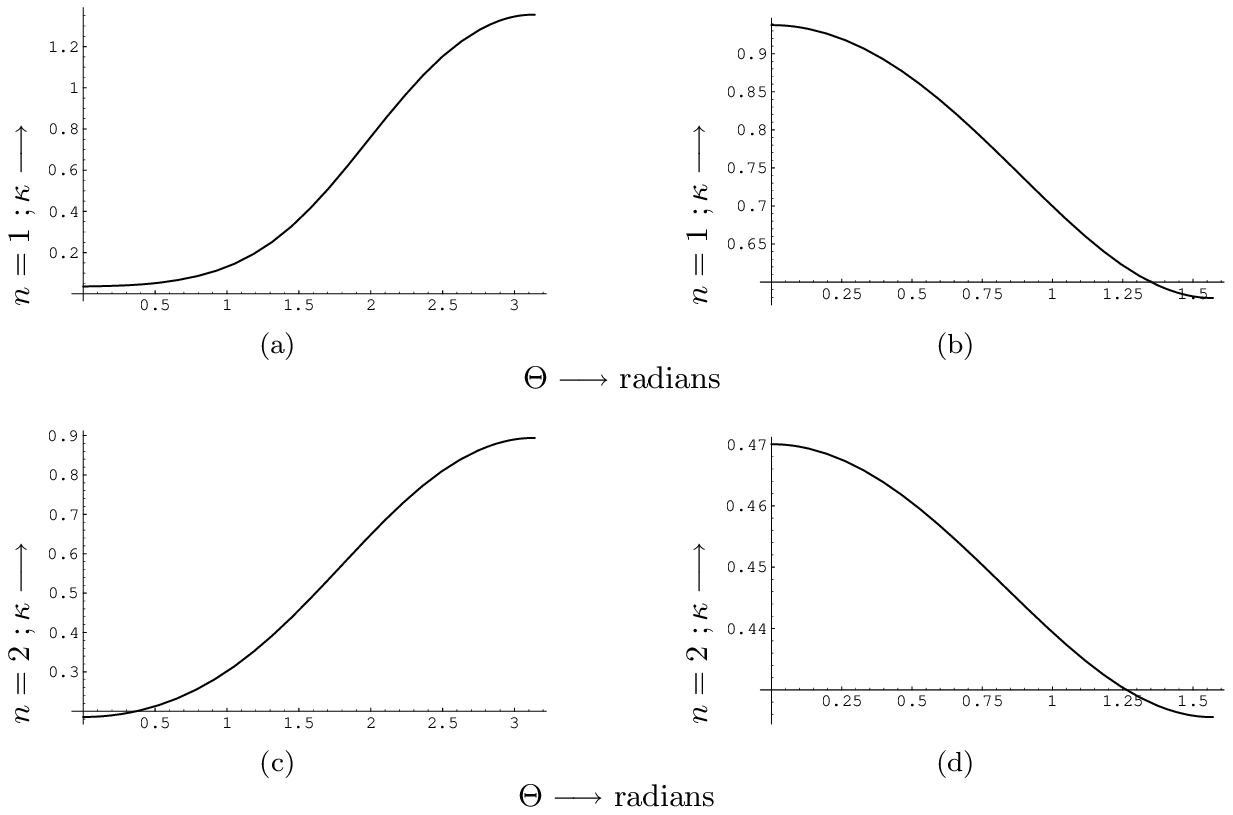}
  \caption{The same as in Fig. \ref{fig:Skappa1} for a WIMP mass
of $30$ GeV.}
 \label{fig:Skappa2}
  \end{center}
  \end{figure}

\begin{figure}[!ht]
 \begin{center}
   \includegraphics[scale=0.8]{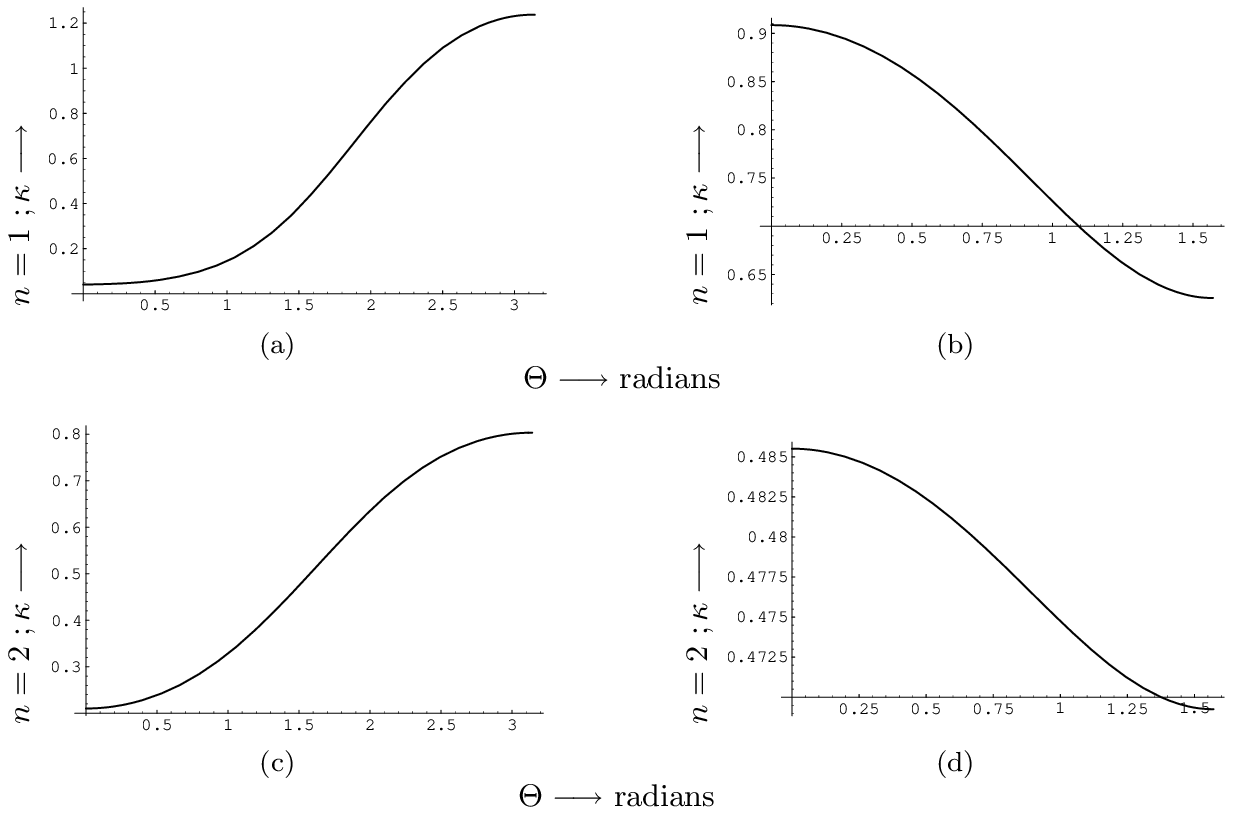}
 \caption{The same as in Fig. \ref{fig:Skappa1} for a WIMP mass
of $100$ GeV.}
 \label{fig:Skappa5}
  \end{center}
  \end{figure}

\begin{figure}[!ht]
 \begin{center}
    \includegraphics[scale=0.8]{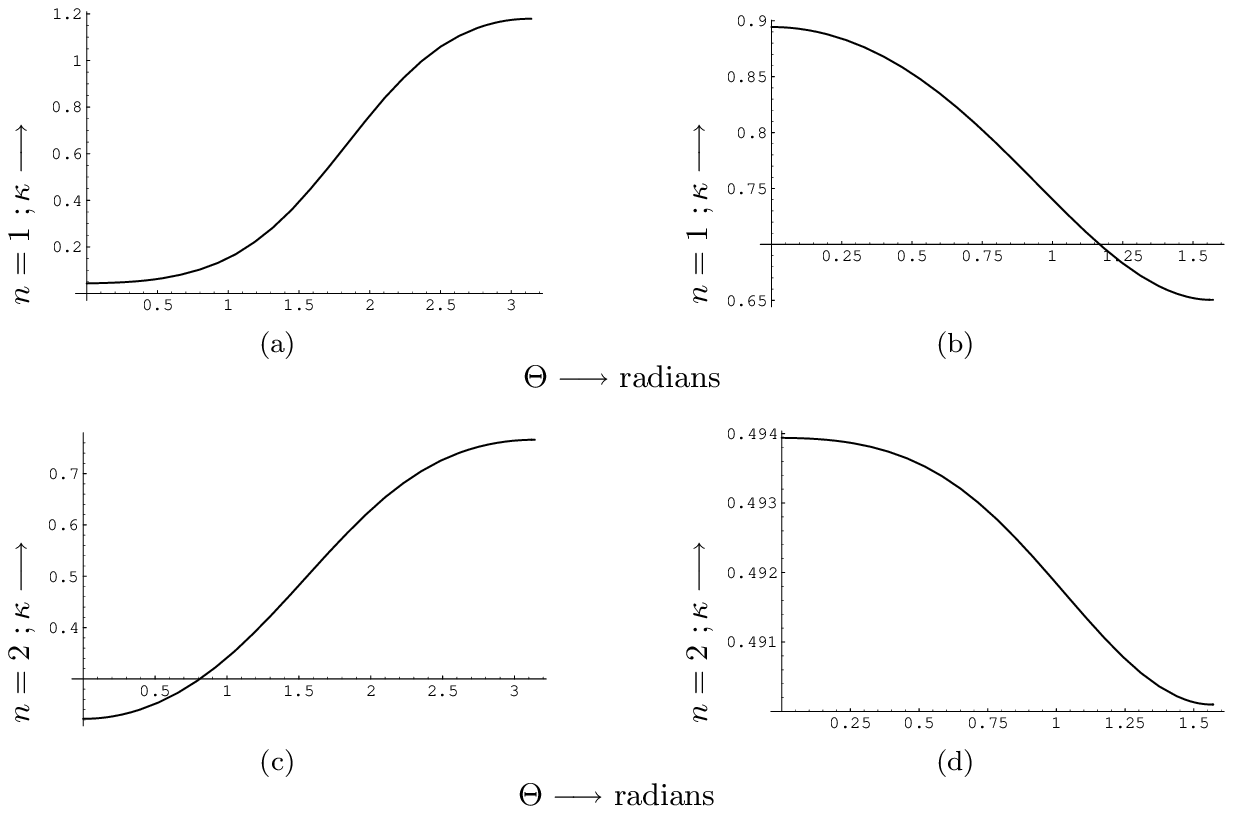}
  \caption{The same as in Fig. \ref{fig:Skappa1} for a WIMP mass
of $250$ GeV.}
 \label{fig:Skappa9}
  \end{center}
  \end{figure}

  We continue our analysis by calculating the parameter $\kappa_{coh}$ discussed above. Our results are
  shown in Figs
  \ref{fig:Skappa1}-\ref{fig:Skappa9}.
From these figures we see that in the case of directional
experiments the maximum value of $\kappa$ is attained when the
observation is made opposite to the sun's direction of motion.
This is expected in the case of the M-B distribution. We have seen
above that $<\xi>$ also attains its maximum value. When the sense
of direction is not observed the maximum event rate is again
attained when the observation is along the line of the sun's
motion.

\begin{figure}[!ht]
 \begin{center}
     \includegraphics[scale=0.8]{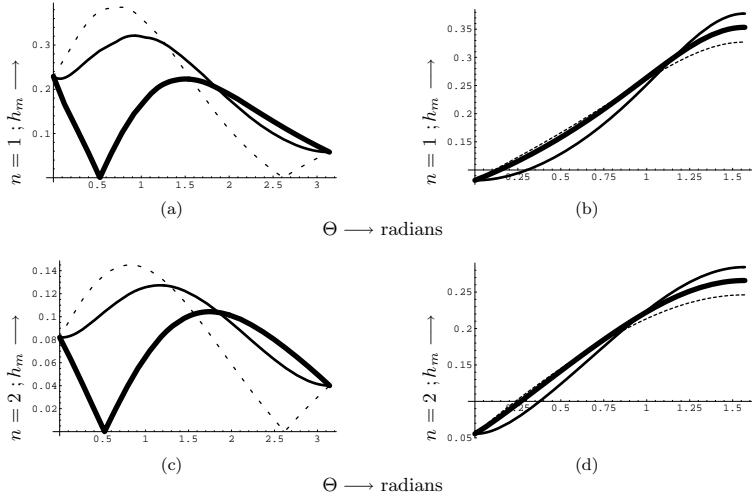}
 \caption{The parameter $h_m$, defined in the text, as function of the polar angle $\Theta$ for a light
target in the case of a WIMP mass of $10$ GeV. The fine solid, thick solid,  dashed and
long dashed curves correspond to $\Phi=0,\pi/2,\pi$, and $ 3\pi/2$
respectively (the cases $\Phi=\pi/2$ and $3\pi/2$ cannot be
distinguished). Note that, when the
sense is not distinguished, the maximum angle of the line of
observation with the sun's direction of motion is $\pi/2$, while
the maximum azimuthal angle is $\pi$. 
 Now the thick solid, fine solid,  dashed and long dashed curve
correspond to $\Phi=0,\pi/4, \pi/2$ and $3 \pi/4$ respectively
(the cases $\Phi=\pi/4$ and $3\pi/4$ cannot be distinguished). (a) The directional rate in a given
sense for $n=1$. (b) The sum of both senses is shown for $n=1$.
 (c) The same as in (a) for $n=2$. (d) the same as in (b) for $n=2$.}
 \label{fig:Shcs1}
  \end{center}
  \end{figure}

\begin{figure}[!ht]
 \begin{center}
   \includegraphics[scale=0.8]{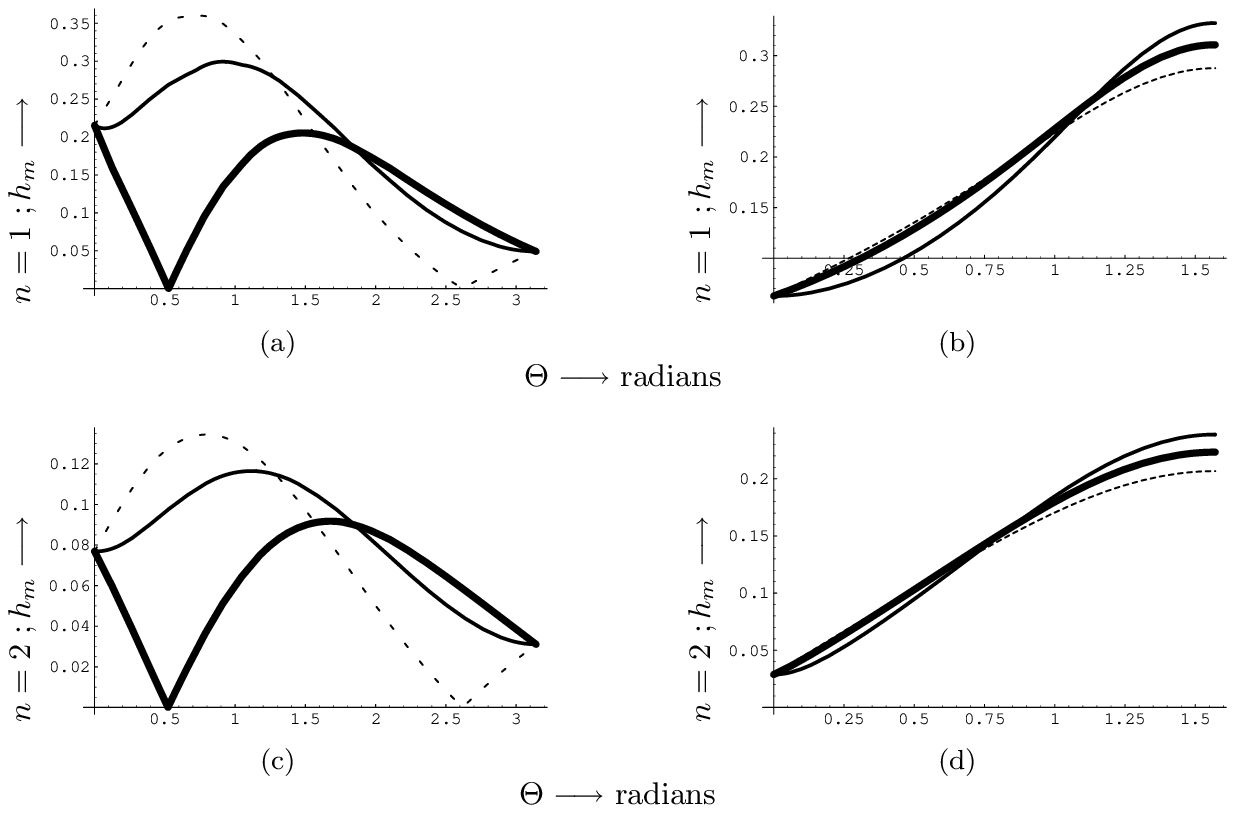}
  \caption{The
same as in Fig. \ref{fig:Shcs1} for a WIMP mass of $30$ GeV.}
 \label{fig:Shcs2}
  \end{center}
  \end{figure}

\begin{figure}[!ht]
 \begin{center}
    \includegraphics[scale=0.8]{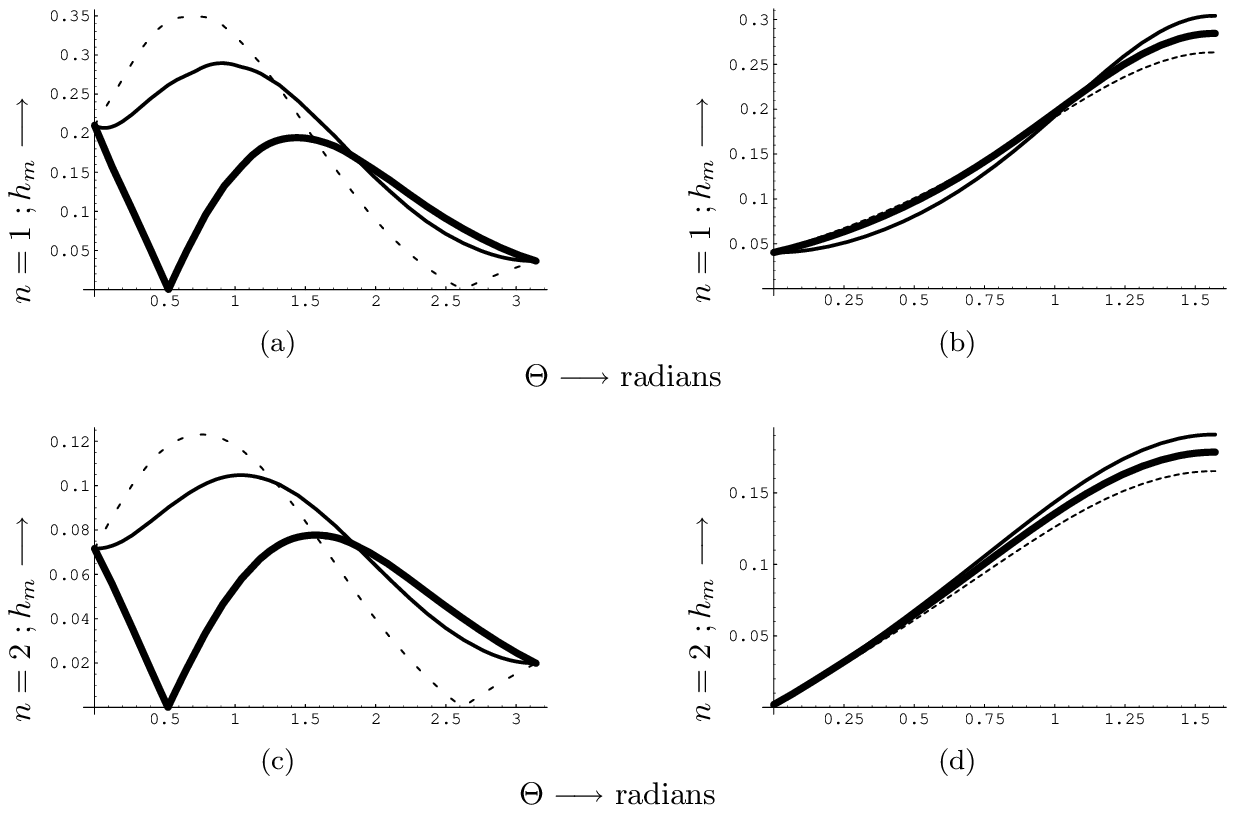}
 \caption{The
same as in Fig. \ref{fig:Shcs1} for a WIMP mass of $100$ GeV.}
 \label{fig:Shcs5}
  \end{center}
  \end{figure}

\begin{figure}[!ht]
 \begin{center}
     \includegraphics[scale=0.8]{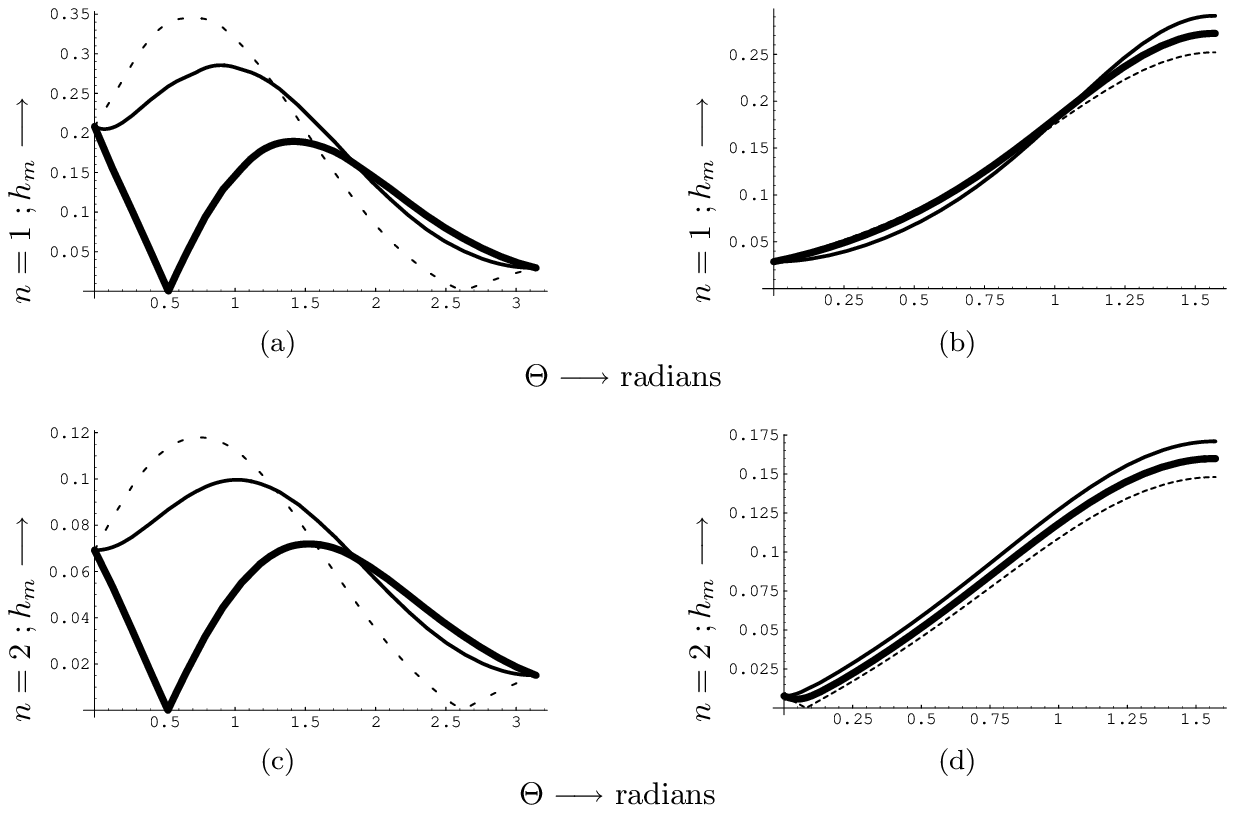}
 \caption{The
same as in Fig. \ref{fig:Shcs1} for a WIMP mass of $250$ GeV.}
 \label{fig:Shcs9}
  \end{center}
  \end{figure}

\begin{figure}[!ht]
 \begin{center}
     \includegraphics[scale=0.8]{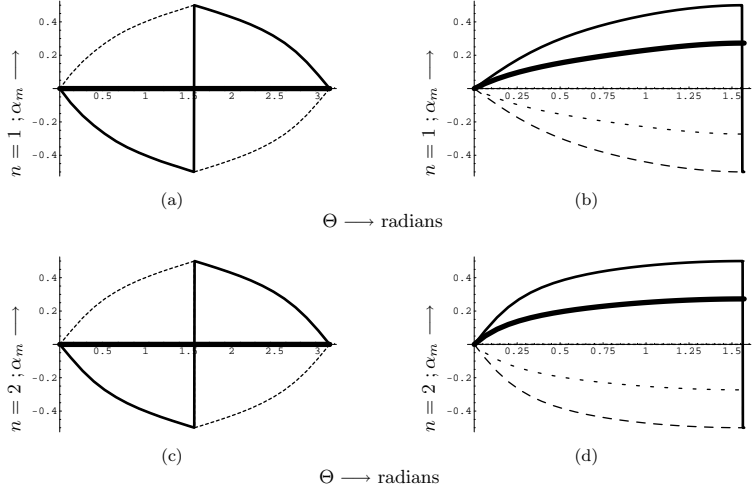}
\caption{The parameter $\alpha_m$ (in units of $\pi$), defined in the text, as function of the polar
 angle $\Theta$ for a light target.
Otherwise the notation is the same as in Fig. \ref{fig:Shcs1}.}
 \label{fig:Sphase1}
  \end{center}
  \end{figure}

\begin{figure}[!ht]
 \begin{center}
     \includegraphics[scale=0.8]{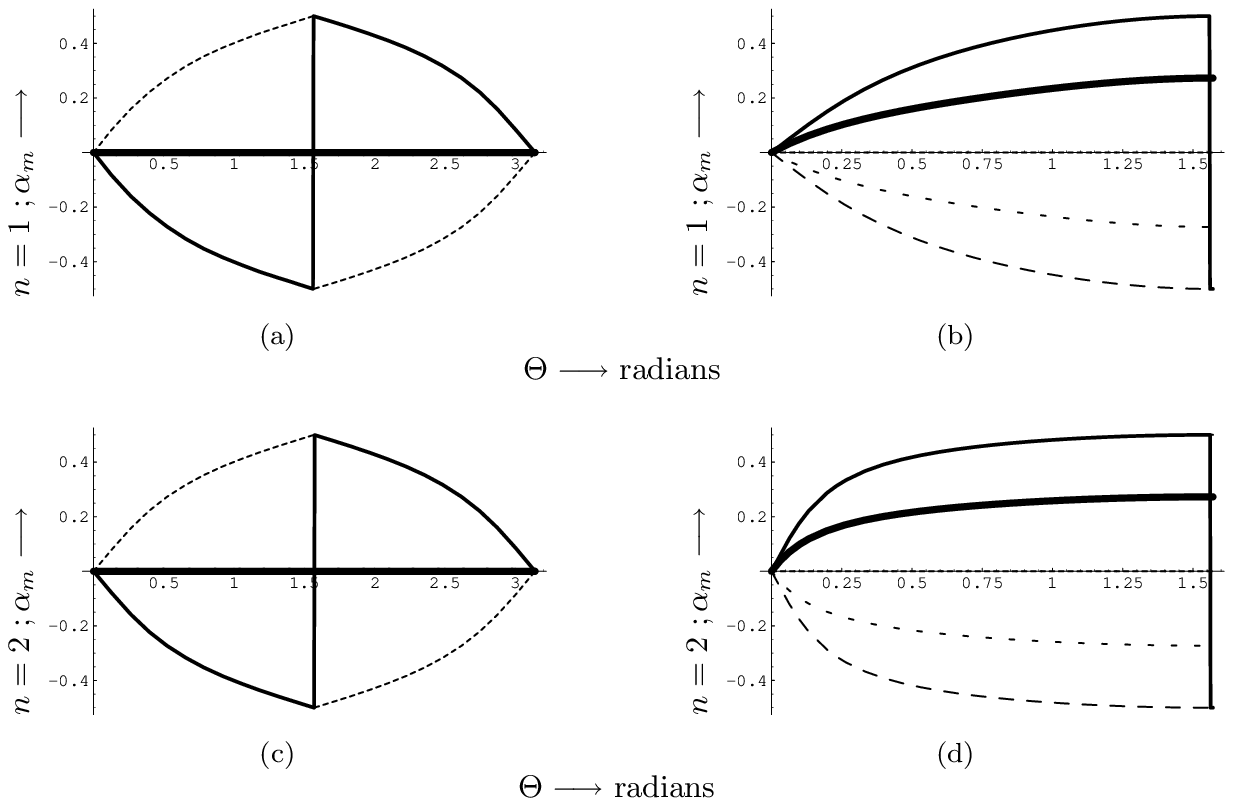}
 \caption{The  same as in Fig. \ref{fig:Sphase1} for a WIMP mass
of $30$ GeV.}
 \label{fig:Sphase2}
  \end{center}
  \end{figure}

\begin{figure}[!ht]
 \begin{center}
     \includegraphics[scale=0.8]{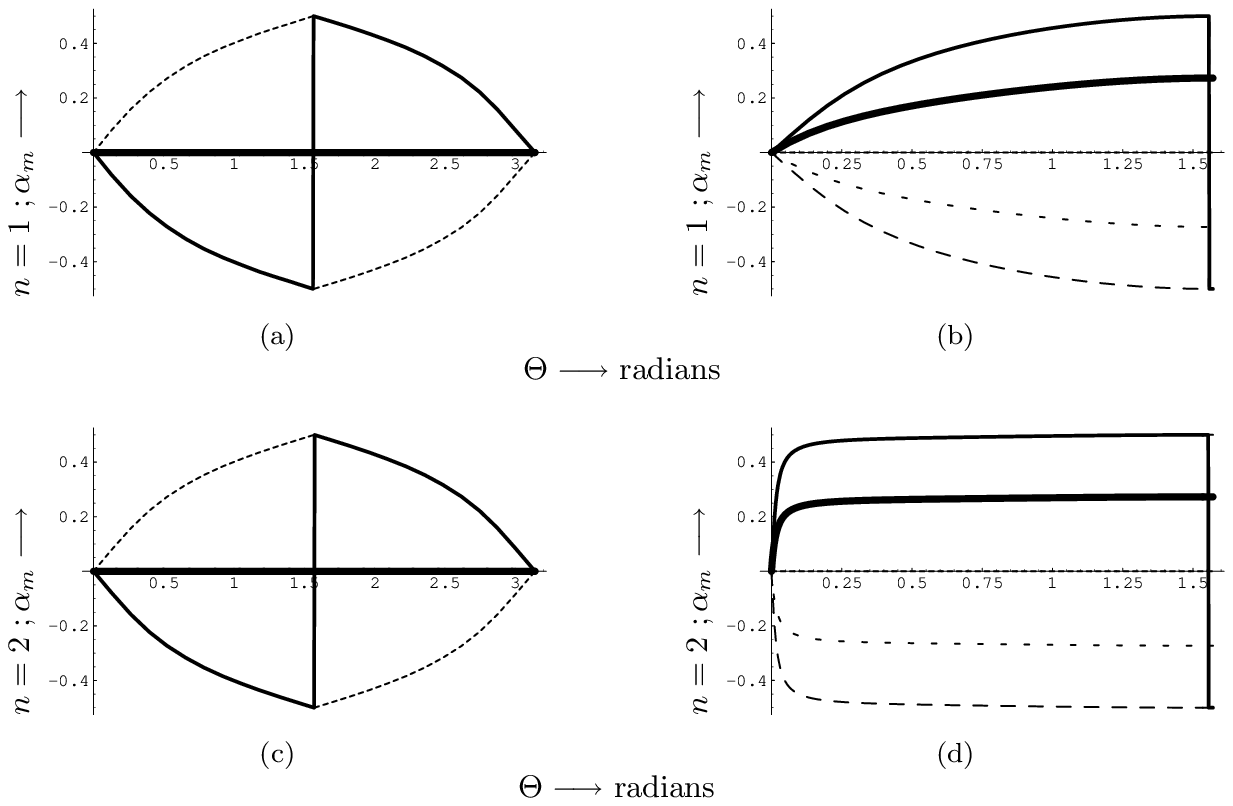}
\caption{ The  same as in Fig. \ref{fig:Sphase1} for a WIMP mass
of $100$ GeV.}
 \label{fig:Sphase5}
  \end{center}
  \end{figure}

\begin{figure}[!ht]
 \begin{center}
      \includegraphics[scale=0.8]{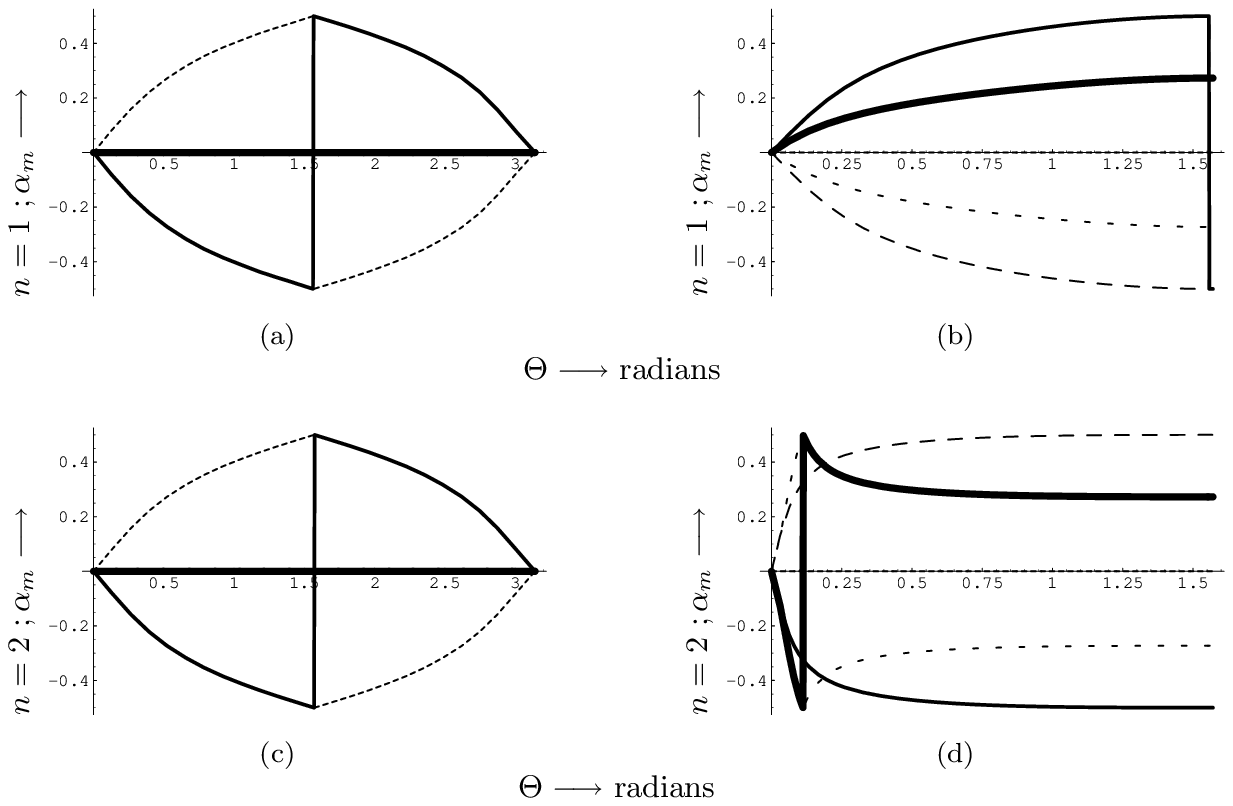}
\caption{ The  same as in Fig. \ref{fig:Sphase1} for a WIMP mass
of $250$ GeV.}
 \label{fig:Sphase9}
  \end{center}
  \end{figure}

 We will next
discus the modulation effect, i.e. the parameters $h_m$,
 shown in Figs \ref{fig:hcs1}-\ref{fig:hcs9} and
$\alpha_m$ shown in Figs \ref{fig:Sphase1}-\ref{fig:Sphase9}. We
clearly see that the results depend on the WIMP mass.
 From the figures
 \ref{fig:Shcs1}-\ref{fig:Shcs9} we see that the modulation for $n=2$ is quite small, reminiscent
 of a similar result in the case of the non directional case \cite{TETRVER06}. Since the exhibited pattern is
 otherwise similar to the standard $n=1$ case we will limit our discussion to the $n=1$ case.
 The modulation amplitude, $h_m\approx0.1$, in the most favored
 direction $\Theta=\pi$, i.e. opposite to the sun's direction of
 motion, is small, but still much larger than that encountered in the non
 directional case.  The absolute maximum $h_m\approx0.4$ seems to occur when $\Phi=\pi$, i.e. inwards towards
  the center of the
  galaxy, at an angle $\Theta=0.5\approx\pi/6$. A smaller value of 0.3 is attained when one is looking radially
   outwards
  from the center of the galaxy at $\Theta=1.0\approx\pi/3$. Another maximum  $h_m\approx0.2$ is attained, when
  $\Theta\approx\pi/2$, i.e. in a plane perpendicular to the sun's direction of
  motion, and $\Phi=\pi/2$, along the line perpendicular to the galactic plane.

  When the sense of direction is not observed the modulation picture is
  quite simple. One sees that the
  modulation is quite small, when the line of observation is along
  the sun's motion and reaches a maximum value, $h_m\approx0.3$ on the plane perpendicular to
  the sun's velocity regardless of $\Phi$.

Regarding the phase $\alpha_m$ we see that, as expected, it is
zero at
 $\Theta=0$ or $\pi$, reminiscent of the non directional case.
 For $\Phi=\pi/2$ or $3 \pi/2$ this phase is zero for all $\Theta$. This means
that in all these cases one has the standard behavior of the
modulation (maximum around June 2nd). For $\Phi=0$, outwards from
the center of the galaxy, the phase decreases from zero to $\pi/2$
depending so long as  $\Theta<0$. During this period the maximum
precedes the phase of the Earth. When $\Theta>\pi/2$, it changes
sign (the maximum drags behind the phase of the Earth). In other
words the maximum occurs in the spring for $\Theta
>\pi/2$ and in the autumn for $\Theta <\pi/2$. The opposite is
true when $\Phi=\pi$.

In the case when both senses of WIMP recoil are considered the
modulation drags behind the phase of the Earth, a maximum in the
spring, for $\Phi=0, \pi$  (radially in the galaxy). It attains a
maximum in the fall for $\phi=p/2, 3 \pi/2$ (perpendicular to the
plane of the galaxy)

 Anyway the above complicated pattern of the shift in the location of the maximum, which depends on the LSP
 mass, is not expected to
  cause problems in the analysis of the experiments. The periodic nature of the phenomenon is sufficient.\\
 \subsection{The intermediate mass target $_{127}$I}
\begin{figure}[!ht]
 \begin{center}
    \includegraphics[scale=0.8]{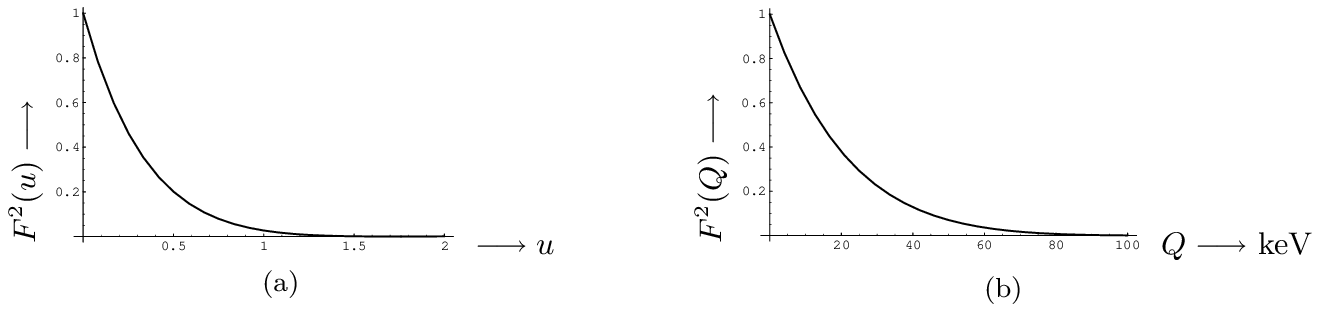}
\caption{(a) The form factor $F^2(u)$ for $^{127}$I employed in our calculation with  $u=Q/Q_0$, Q  the energy transfer to the nucleus   and $Q_0=64$ keV. (b) The same quantity a function of the energy transfer $Q$.}
 \label{sqformf127}
   \end{center}
  \end{figure}

\begin{figure}[!ht]
 \begin{center}
    \includegraphics[scale=0.8]{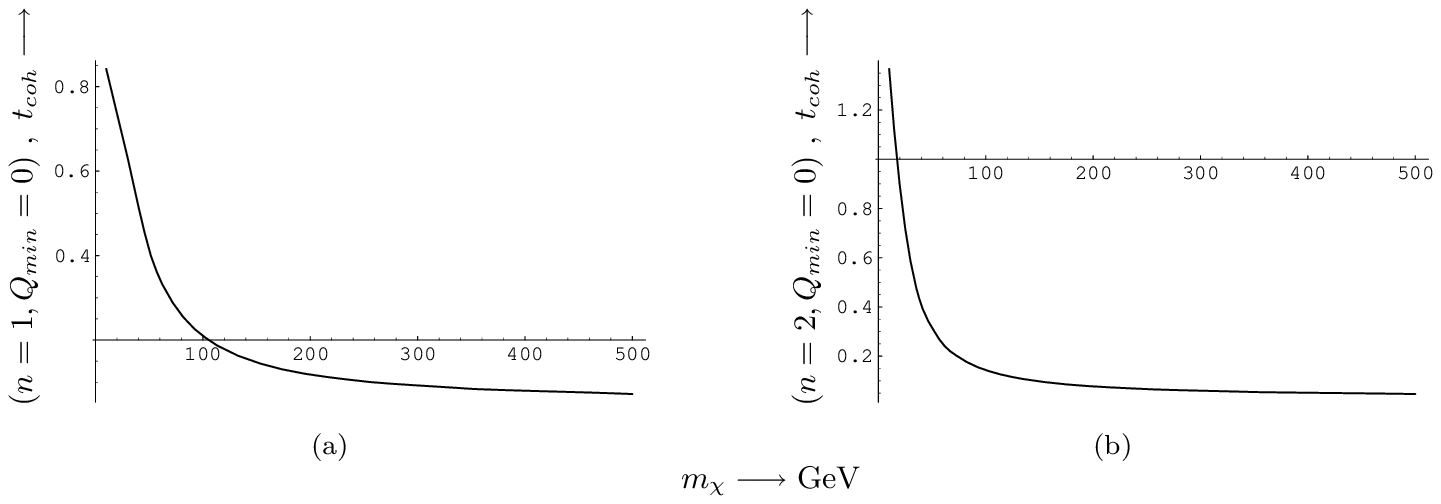}
 \caption{The quantity $t_{coh}$ is shown for $Q_{min}=0$ with (a) $n=1$ and (b) $n=2$.}
 \label{fig:totalt}
   \end{center}
  \end{figure}

\begin{figure}[!ht]
 \begin{center}
   \includegraphics[scale=0.8]{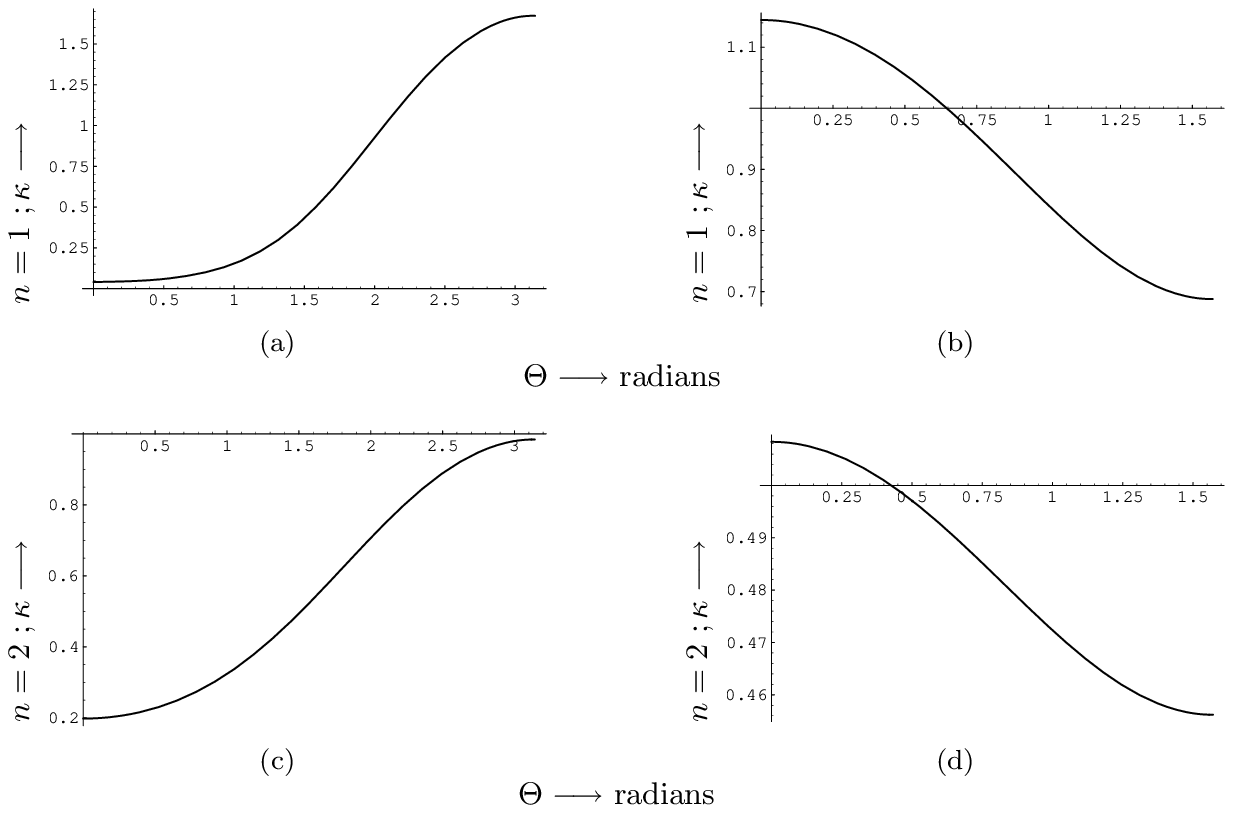}
 \caption{The same as in Fig. \ref{fig:Skappa1} for the target $^{132}$I.}
 \label{fig:kappa1}
  \end{center}
  \end{figure}

\begin{figure}[!ht]
 \begin{center}
   \includegraphics[scale=0.8]{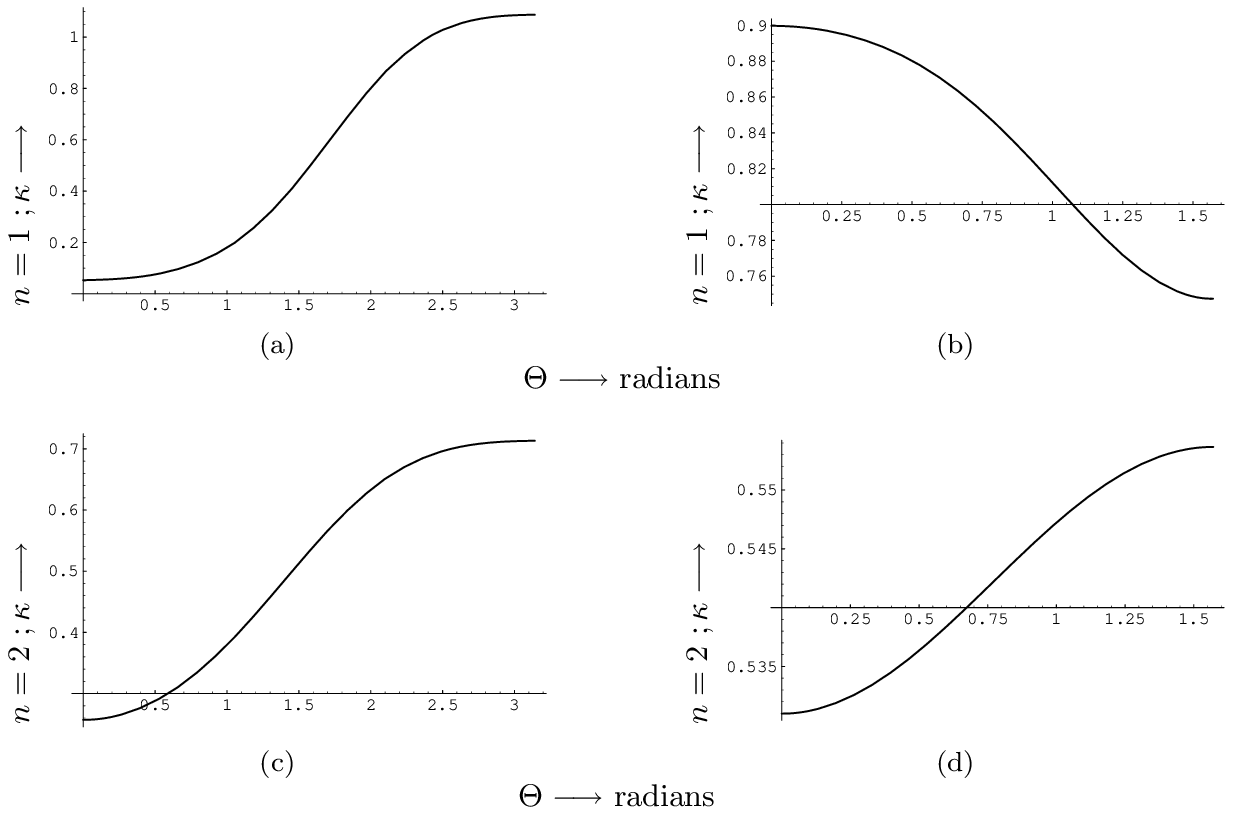}
\caption{The same as in Fig. \ref{fig:kappa1} for a WIMP mass
of $30$ GeV.}
 \label{fig:kappa2}
  \end{center}
  \end{figure}

\begin{figure}[!ht]
 \begin{center}
    \includegraphics[scale=0.8]{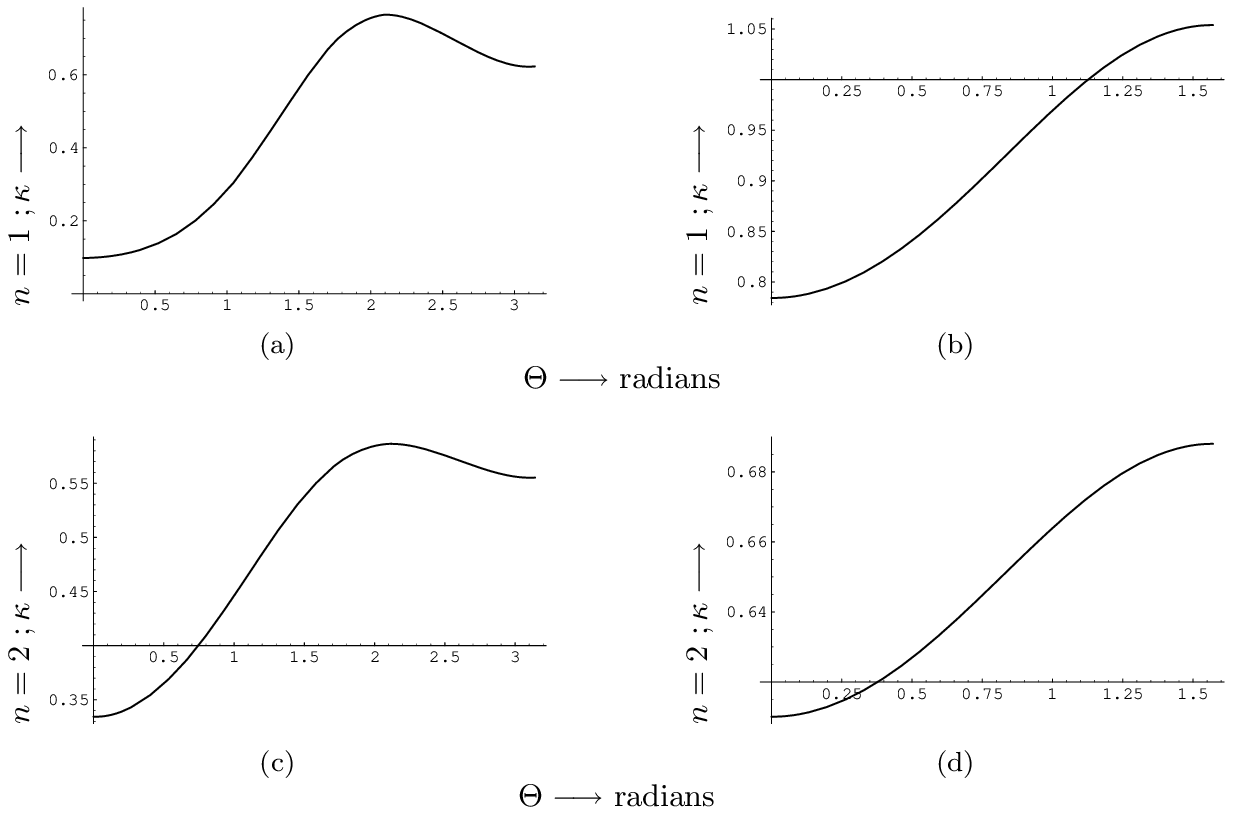}
\caption{The same as in Fig. \ref{fig:kappa1} for a WIMP mass
of $100$ GeV.}
 \label{fig:kappa5}
  \end{center}
  \end{figure}

\begin{figure}[!ht]
 \begin{center}
     \includegraphics[scale=0.8]{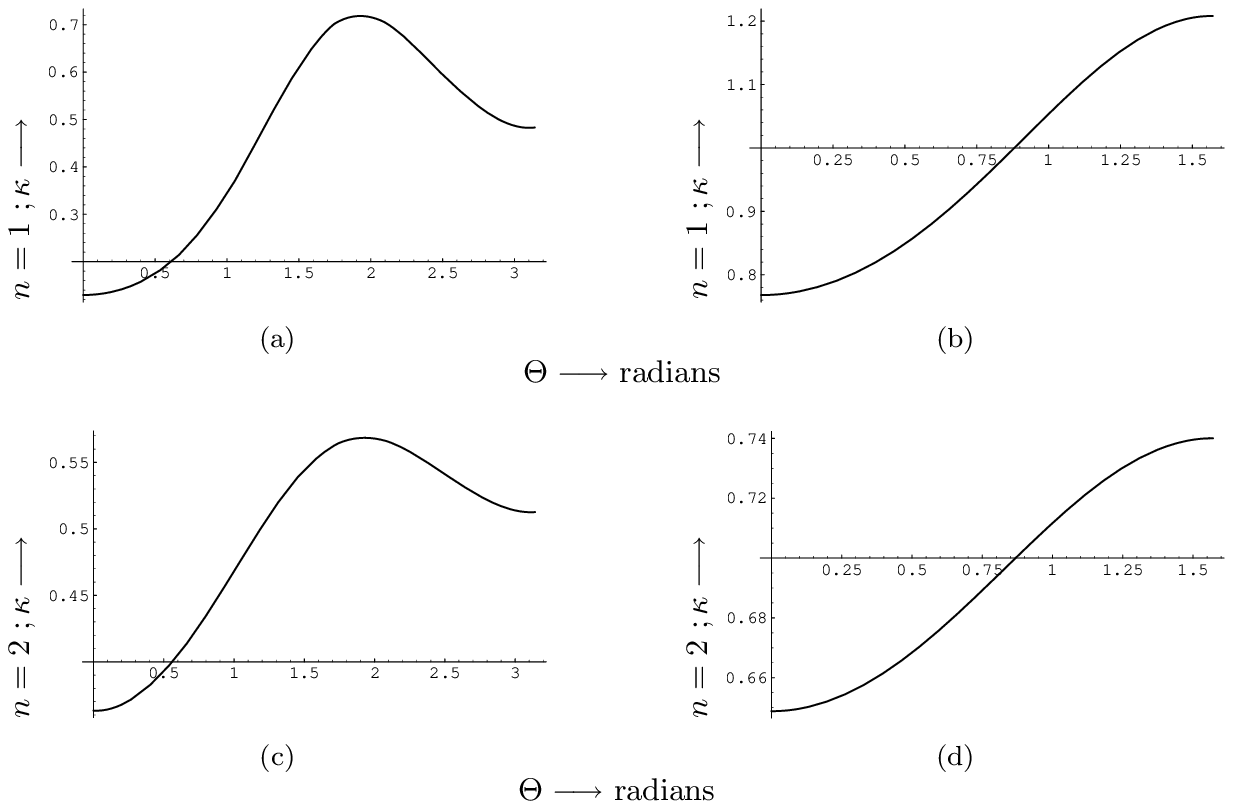}
 \caption{The same as in Fig. \ref{fig:kappa1} for a WIMP mass
of $250$ GeV.}
 \label{fig:kappa9}
  \end{center}
  \end{figure}

The nuclear form factor employed was obtained in the shell model description of the target and is shown in
 Fig. \ref{sqformf127}.
 We first exhibit the previously obtained \cite{TETRVER06} parameter $t_{coh}$ in Fig. \ref{fig:totalt}. We have a
  suppression as the WIMP mass increases (partly counteracted by the factor $\mu_r^2/m_{WIMP}$ entering the event
   rate and not included in $t_{coh}$).
 The behavior of the parameter $\kappa$ is exhibited in Figs \ref{fig:kappa1}-\ref{fig:kappa9}. One can see that
 for relatively light WIMP  in the directional case $\kappa$ has a similar behavior with that encountered
  above for a light
 target. It peaks in the direction opposite to the sun's velocity,
i.e at $\Theta=\pi$, as expected. For heavy WIMP masses the
  maximum value is a factor of 2 smaller than that for a light target and it occurs
at lower polar angles $\Theta$. When both senses are
 considered for $n=1$ $\kappa$ peaks along the sun's line of motion (by definition here $\Theta=0$).
 The situation is reversed for heavy WIMP, i.e. it attains a maximum at $\Theta=\pi/2$. This reversed pattern holds
 for $n=2$, except for very small WIMP masses.  Thus,
if one could independently establish that the velocity distribution is of M-B type, this signature can be used in
constraining the WIMP mass.

\begin{figure}[!ht]
 \begin{center}
     \includegraphics[scale=0.8]{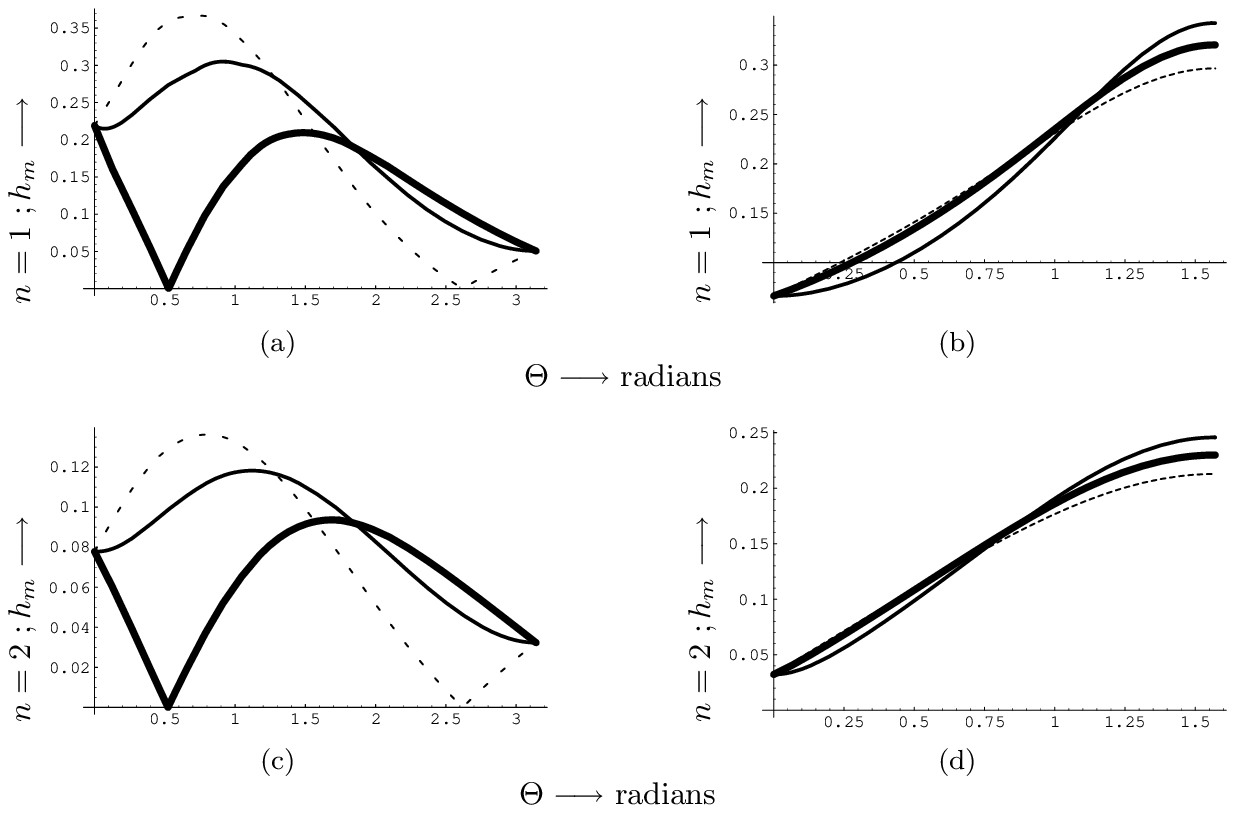}
 \caption{The same a in Fig. \ref{fig:Shcs1} for the target
$^{127}$I.}
 \label{fig:hcs1}
  \end{center}
  \end{figure}

\begin{figure}[!ht]
 \begin{center}
      \includegraphics[scale=0.8]{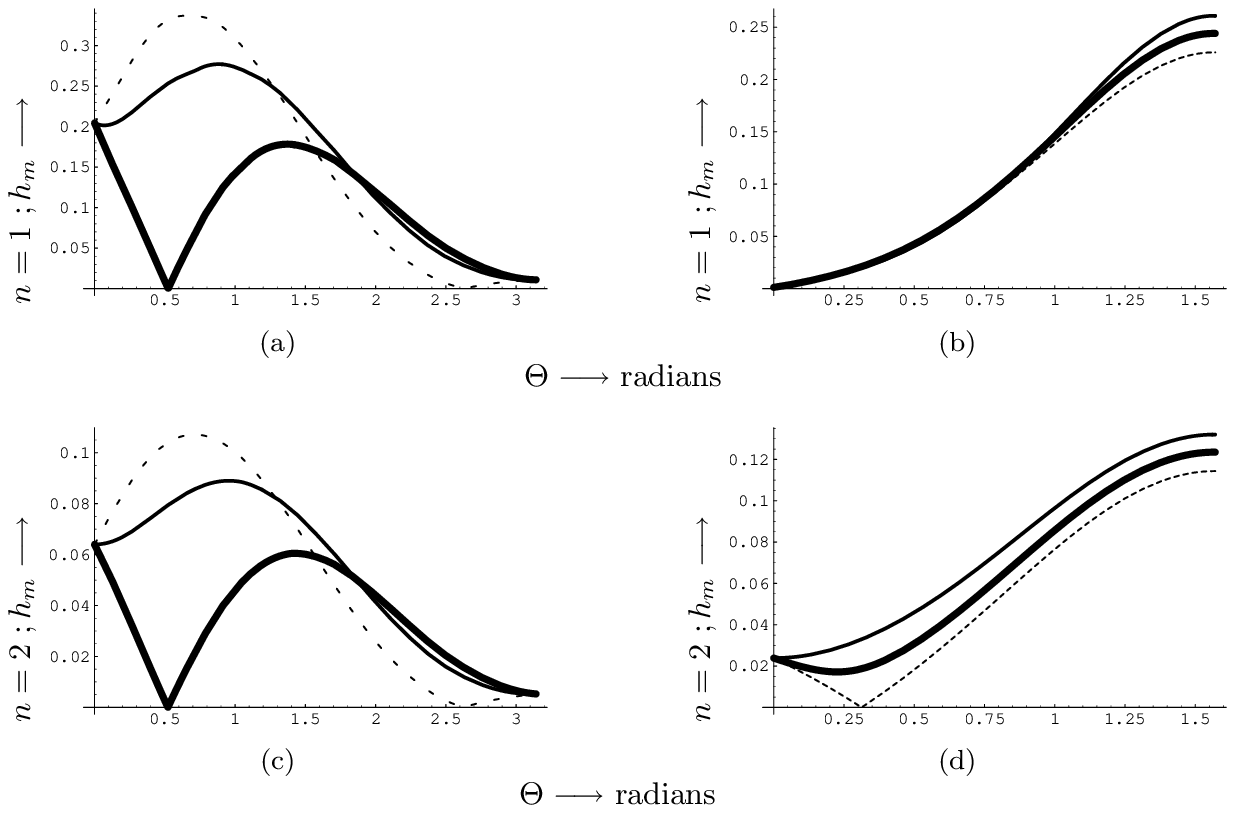}
 \caption{The
same as in Fig. \ref{fig:hcs1} for a WIMP mass of $30$ GeV.}
 \label{fig:hcs2}
  \end{center}
  \end{figure}

\begin{figure}[!ht]
 \begin{center}
    \includegraphics[scale=0.8]{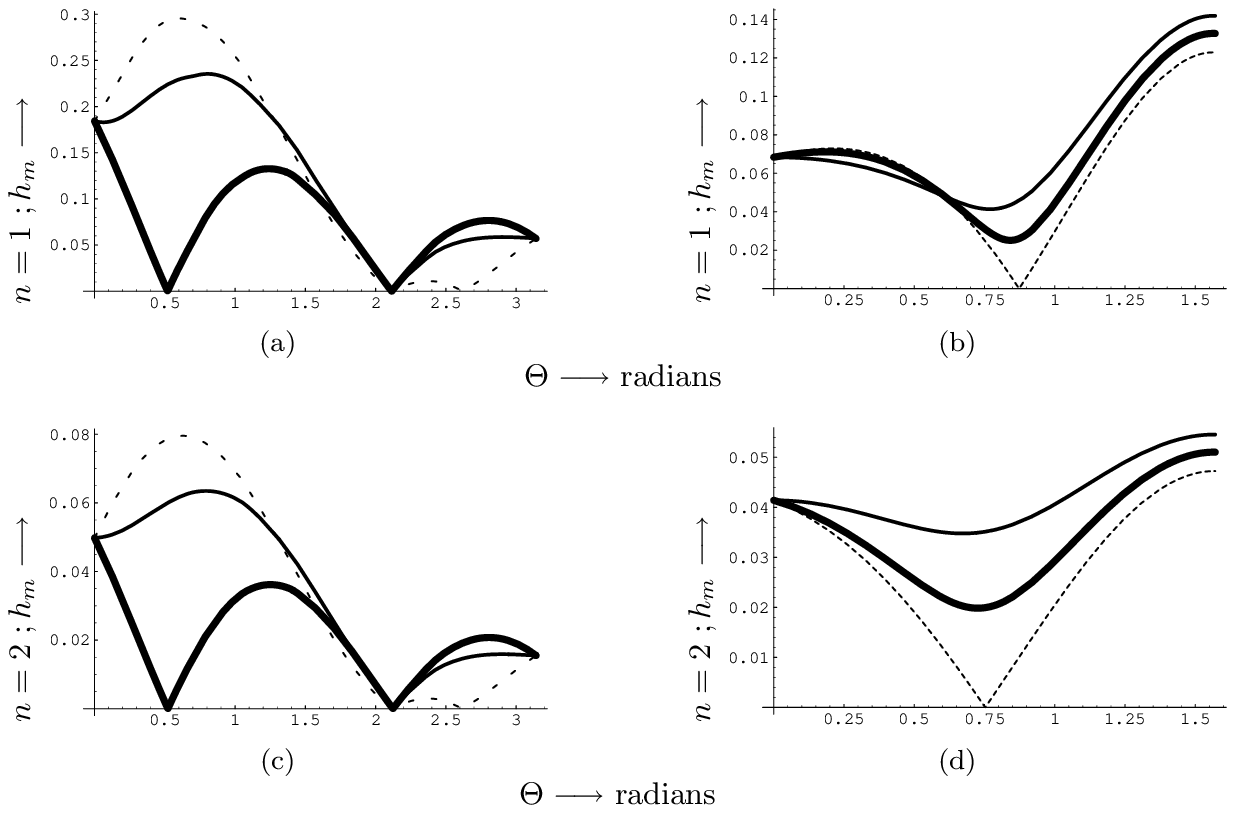}
  \caption{The
same as in Fig. \ref{fig:hcs1} for a WIMP mass of $100$ GeV.}
 \label{fig:hcs5}
  \end{center}
  \end{figure}

\begin{figure}[!ht]
 \begin{center}
     \includegraphics[scale=0.8]{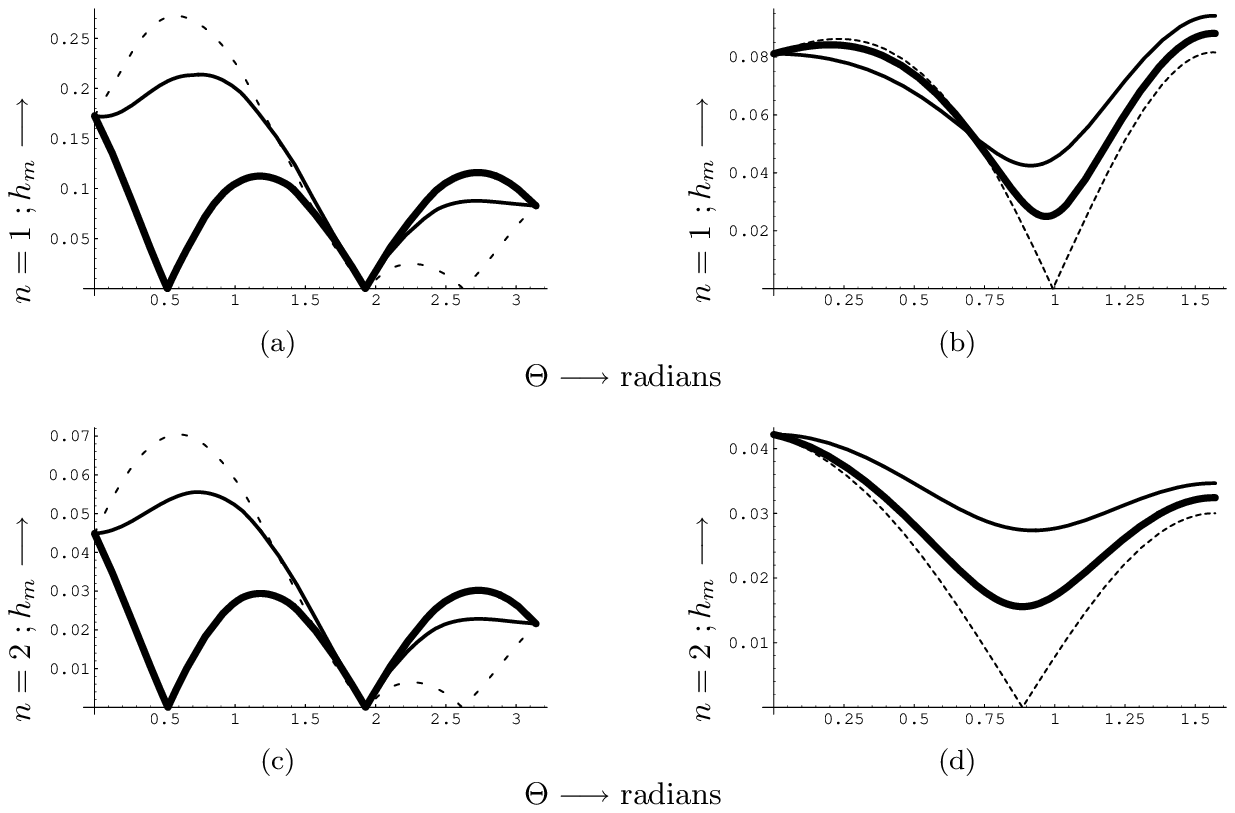}
 \caption{The
same as in Fig. \ref{fig:hcs1} for a WIMP mass of $250$ GeV.}
 \label{fig:hcs9}
  \end{center}
  \end{figure}

We will next discus the modulation effect, i.e. the parameters $h_m$,
 shown in Figs \ref{fig:hcs1}-\ref{fig:hcs9}.
 We
clearly see that the results depend on the WIMP mass. From the
figures
 \ref{fig:hcs1}-\ref{fig:hcs9} we see that in the directional case the modulation is similar
 to that for a light target for small WIMP masses. This is not surprising, since, then, the reduced
 mass the same for both targets. For heavy WIMPs the modulation $h_m$ goes through a small value at around
 $\Theta=(2/3)\pi$ regardless of the angle $\Phi$. The maximum value of $h_m$ is also a bit smaller.

 If both senses are counted the modulation
 amplitude becomes essentially
independent of $\Phi$ and attains the maximum value of
$h_m=0.1-0.3$ depending on the WIMP mass. Note that in this case
for heavy WIMPs one has a secondary maximum and a minimum at
$\Theta\approx(3/4)\approx\pi/4$, .i.e. when the recoil occurs in
the plane perpendicular to the sun's velocity. Again the absolute
maximum occurs at $\Theta=\pi$, i.e. along the sun's motion.

We will not show the phase $\alpha_m$, since the obtained pattern is similar to that
encountered for a light target.
  Before concluding this section we should mention again that in the case of $^{127}$I one may have a contribution
  due to the spin cross section. As we have already mentioned, the controversy regarding the DAMA experiment, 
employing NaI target, may be
attributed to this interaction, which does not enter in experiments involving even nuclear targets.
This possibility is currently under study, including  our own realistic (static) spin ME and spin 
form factors. The quantities $t,h,dr/d \xi,\kappa$ and $h_m$ do not depend on the spin ME, they only 
depend on the adopted
spin form factors. We do not, however, expect the factors $dr/d \xi,\kappa$ and $h_m$, which are the main subject
of this work, to be radically different from those presented  here for the coherent cross section.

\section{Conclusions}
 We have seen that, given a sufficient number of events, the directional experiments, in which one measures
 the direction of the nuclear recoil,  provide an excellent signature to discriminate against background.
 Some of these features persist even if the sense of motion of recoils along their line of motion cannot
 be measured. The predictions depend, of course, on the assumed velocity distribution. In the present work
 we selected to work with a M-B distribution: (i) the traditional one with characteristic velocity that of
 sun around the center of the galaxy and (ii) a variant obtained when dark matter is coupled to dark energy
 via a scalar field yielding an increase in the gravitational field for dark matter \cite{TETRVER06}.
  In the latter case the
  characteristic velocity in the WIMP distribution increases by a factor                                                                                                                                                    or
  $n\geq1$.

In the most favored direction, opposite to the sun's direction of
motion, the event rate is $\approx \frac{1}{2 \pi}$ down from the
standard non directional experiments.  The modulation amplitude in
this direction
 $h_m$ depends on the WIMP mass. For a light target it ranges between 0.05 and and 0.1 depending on the
 WIMP mass. For a heavy target it is somewhat reduced, $h_m=0.02-0.07$, but it remains higher than that
 expected in the non directional experiments. It is also characterized by a definite sign (maximum
 around June 3nd). Higher values $h_m=0.25-0.35$, yielding $50-70\%$ difference between the
 maximum and the minimum rates, are expected by a judicious choice of the
 direction of observation. Thus quite large asymmetries with seasonal dependence are predicted.
 The time  of the maximum is also direction dependent, so it cannot
 be mimicked by seasonal variations of the background.

 In partly directional experiments, i.e. experiments in which the sense of motion of the recoiling
 nucleus is not determined, one no longer can measure
 asymmetries. We still find, however, that the expected event rate
 is maximum along the line of motion. Equally large modulation
 amplitudes $h_m$ are predicted with a seasonal variation, which,
 again, cannot be mimicked by seasonal variations in the
 background.

The modulation amplitude is decreased, if dark matter is coupled with dark energy, in a fashion analogous to the
 non directional case. This is not unexpected, since the ratio of the velocity of the earth to the characteristic
  velocity becomes smaller. This suggests that one should test whether the above conclusions hold,
  by considering other velocity distributions. Thus one may be able
to infer the velocity distribution, when
 the experimental data become available.

 The directional experiments are quite hard. It is encouraging that the angular distribution is such that
 the average angle is given by
 $<\xi>=<\cos{\theta}>=0.5-0.8$ depending on the angle $\Theta$, the polar angle of the direction of
 observation from the sun's direction of motion. Anyway in experiments
 planned by DRIFT \cite{DRIFT} one can register events recoiling in all directions.
 If some interesting events are
 found, one may further analyze them by the direction of recoil to discriminate against background.
  \section{Acknowledgments:} The work of one of the authors (JDV) was completed during a visit to Tuebingen
   as a Humboldt Research Awardee.

\end{document}